\documentclass[prb,showpacs,twocolumn,floats,superscriptaddress,nolongbibliography,10pt]{revtex4-2}
\usepackage[pdftex]{graphicx}
\usepackage{amsmath,amssymb,euscript,wasysym}
\usepackage{cases} 
\usepackage{xspace}

\def\imag{\mathrm{Im}\,} 
\def\real{\mathrm{Re}\,} 

\def\i{\textnormal{i}}

\def\Jk{\ensuremath{J_{\mathrm{K}}}}
\def\Tk{\ensuremath{T_{\mathrm{K}}}}
\def\TspinO{\ensuremath{T_{\spin}^{\mathrm{onset}}}}
\def\TspinC{\ensuremath{T_{\spin}^{\mathrm{cmp}}}}
\def\Tspin{\ensuremath{T_{\spin}}}

\def\TorbO{\ensuremath{T_{\orb}^{\mathrm{onset}}}}
\def\TorbC{\ensuremath{T_{\orb}^{\mathrm{cmp}}}}
\def\Torb{\ensuremath{T_{\mathrm{\orb}}}}

\def\TM{\ensuremath{T_{\mathrm{M}}}}
  
\newcommand{\vo}{V$_2$O$_3$}

\newcommand{\sroone}{Sr$_2$RuO$_4$}

\newcommand{\HHM}{{3HHM}\xspace}

\newcommand{\imp}{{\textrm{imp}}}

\newcommand{\interact}{{\textrm{int}}}
\newcommand{\charge}{{\textrm{ch}}}
\newcommand{\CW}{{\scriptscriptstyle {\textrm{CW}}}}
\newcommand{\spin}{{\textrm{spin}}}
\newcommand{\orb}{{\textrm{orb}}}

\newcommand{\pdag}{{\phantom{\dagger}}}

\newcommand{\latt}{{\textrm{latt}}}

\newcommand{\eff}{{\textrm{{\textrm{eff}}}}}
\newcommand{\fl}{\textrm{FL}}
\newcommand{\cross}{\textrm{cr}}
\newcommand{\ARPESt}{ARPES}
\newcommand{\MIR}{\textrm{MIR}}

\newcommand{\cmp}{\textrm{cmp}}

\newcommand{\Mone}{\textrm{M1}}
\newcommand{\Hone}{\textrm{H1}}

\newcommand{\Itwo}{\textrm{I2}}
\newcommand{\Wzero}{\textrm{W0}}

\newcommand{\chispin}{\chi_{\rm{spin}}}
\newcommand{\chiorb}{\chi_{\rm{orb}}}
\newcommand{\chizero}{\chi_{0}}

\newcommand{\Sec}[1]{Sec.~\ref{#1}}
\newcommand{\App}[1]{Appendix~\ref{#1}}
\newcommand{\Eq}[1]{Eq.~\eqref{#1}}
\newcommand{\Eqs}[1]{Eqs.~\eqref{#1}}

\newcommand{\Fig}[1]{Fig.~\ref{#1}}
\newcommand{\Figs}[1]{Figs.~\ref{#1}}

\newcommand{\oRef}[1]{Ref.~[\onlinecite{#1}]}

\RequirePackage[
   hyperindex,colorlinks,bookmarksnumbered,
  plainpages=true,pdfstartview=FitH]{hyperref}
 \hypersetup{linkcolor=blue,urlcolor=blue,citecolor=blue}
\usepackage{hyperref}

\begin{document}

\title{Differentiating Hund from Mott physics in a three-band Hubbard-Hund model: Temperature dependence of spectral, transport, and thermodynamic properties}
\author{K. M. Stadler} 
\affiliation{Arnold Sommerfeld Center for Theoretical Physics, 
Center for NanoScience,\looseness=-1\,  and Munich 
Center for \\ Quantum Science and Technology,\looseness=-2\, Ludwig-Maximilians-Universität München, 80333 Munich, Germany}
\author{G. Kotliar} 
\email{kotliar@physics.rutgers.edu}
\affiliation{Department of Physics and Astronomy, 
Rutgers University, Piscataway, NJ 08854, USA} 
\author{S.-S. B. Lee} 
\affiliation{Arnold Sommerfeld Center for Theoretical Physics, 
Center for NanoScience,\looseness=-1\,  and Munich 
Center for \\ Quantum Science and Technology,\looseness=-2\, Ludwig-Maximilians-Universität München, 80333 Munich, Germany}
\author{A. Weichselbaum} 
\affiliation{Condensed Matter Physics and Materials Science Department,
Brookhaven National Laboratory, Upton, New York 11973, USA}
\affiliation{Arnold Sommerfeld Center for Theoretical Physics, 
Center for NanoScience,\looseness=-1\,  and Munich 
Center for \\ Quantum Science and Technology,\looseness=-2\, Ludwig-Maximilians-Universität München, 80333 Munich, Germany}
\author{J. von Delft} 
\email{Correspondence: vondelft@lmu.de}
\affiliation{Arnold Sommerfeld Center for Theoretical Physics, 
Center for NanoScience,\looseness=-1\,  and Munich 
Center for \\ Quantum Science and Technology,\looseness=-2\, Ludwig-Maximilians-Universität München, 80333 Munich, Germany}

\begin{abstract}
We study the interplay between Mott physics, driven by Coulomb repulsion $U$, and Hund physics, driven by Hund's coupling $J$,
 for a minimal model for Hund metals, the orbital-symmetric three-band Hubbard-Hund model (\HHM) for a lattice filling of $1/3$.   Hund-correlated metals are characterized 
by spin-orbital separation (SOS), a Hund's-rule-induced  two-stage Kondo-type screening process, in which spin screening occurs at much lower energy scales than orbital screening. By contrast, in Mott-correlated metals, 
lying close to the phase boundary of a metal-insulator transition, the SOS window becomes negligibly small and the Hubbard bands are well separated. 
Using dynamical mean-field theory and the numerical renormalization group
as real-frequency impurity solver, we identify numerous fingerprints 
distinguishing Hundness from Mottness in the temperature dependence
of various physical quantities. These include ARPES-type spectra, the local self-energy, static local orbital and spin susceptibilities, resistivity, thermopower, and lattice and impurity entropies. Our detailed description of the behavior of these quantities within the context of a simple model Hamiltonian will be helpful for distinguishing Hundness from Mottness in experimental and theoretical studies of real materials.
\end{abstract}
 
\maketitle

\section{Introduction} 
\label{sec:Intro}

The properties of multiorbital metals with strong onsite atomic-like interactions is governed by strong correlation effects.
In this paper, we study the interplay of 
two distinct manifestations of local interactions:  ``Mott physics'', driven by the Coulomb repulsion $U$ governing charge dynamics; 
and ``Hund physics'',  driven by the Hund's rule coupling $J$ affecting spin dynamics.

For many years, strong electronic correlations in metals have  mainly  been associated with Mottness, well-known from ordinary Mott-Hubbard systems --  in the proximity of a Mott-insulating state, $U$ is large 
(compared to $J$) and slows down or even suppresses the electronic
motion. 
This leads to characteristic spectral signatures  like well-separated Hubbard sidebands and fairly flat bands at the Fermi level at low energies and
temperatures, reflecting strongly renormalized heavy Landau
quasiparticles (QPs). At high energies, typically, the quasiparticle
band vanishes and a gap or pseudogap opens between the Hubbard
sidebands.
A well-known example is \vo\ \cite{McWhan1969,McWhan1973,McWhan1973a,Georges1996,Lee2006,Deng2014,Deng2019}.  

Starting around 2008, it has been recognized 
that noticeable correlation effects are manifest in many multiorbital systems far from a Mott insulating state as they have occupancies differing from half integer filling 
\cite{Werner2008,Haule2009,Yin2011a,Yin2011b,Lanata2013,Fanfarillo2017,Kostin2018,Huang2020,Moon2020,Watzenboeck2020,Gorni2021,VillarArribi2021,deMedici2011,Yin2012,Mravlje2011,Dang2015,Kugler2019a,Zingl2019,Linden2020,Dasari2016,HJLee2020,Clepkens2021,Georges2013,Bascones2016,deMedici2017a,Stadler2015,Aron2015,Stadler2018,Kugler2019,Deng2019,Wang2020,Horvat2019,Walter2020,Karp2020,Hoshino2015,Mravlje2016,Wang2020a,Kang2020,Rincon2014,Belozerov2018,Mezio2019,Facio2019,Steinbauer2019,Song2020,Coleman2020,Chatzieleftheriou2020,HJLee2021}. 
In these systems the effect of $U$ is considered to be too small to correlate the electrons, while  Hund's coupling $J$
is only slightly smaller in the solid than for a bare atom~\cite{Marel1988}. These so-called Hund metals are multiorbital systems with rather broad bands and thus sizeable $J$ compared to a strongly screened $U$. 
By now the 3d iron-based superconductors \cite{Haule2009,Yin2011a,Yin2011b,Lanata2013,Fanfarillo2017,Kostin2018,Huang2020,Moon2020,Watzenboeck2020,Gorni2021,VillarArribi2021} and the 4d-based ruthenates \cite{Werner2008,Mravlje2011,Dang2015,Dasari2016,Zingl2019,Kugler2019a,Linden2020,HJLee2020,Clepkens2021} have been studied from this perspective.
Other examples where Hund-rule physics is important are 
iron impurities on a platinum surface~\cite{Khajetoorians2015}, weak itinerant ferromagnets~\cite{Chen2020}, $e_g$ systems such as NiS${}_{2-x}$Se${}_x$~\cite{Jang2021}, the recently discovered Ni-based superconductors~\cite{Wang2020a,Kang2020}, and even cold atom systems~\cite{Richaud2021}.
For some early reviews, see Refs.~\cite{Georges2013,Bascones2016,deMedici2017a}.

Hund metals have many unusual characteristics, including
the following: (i) Atomic histograms showing the 
probability weight for different electronic configurations
are broad. A range of configurations featuring different orbital occupancies all receive significant weight  (implying metallic behavior),
and high-spin multiplets are favored (thus allowing for a quasi-localized spin)~\cite{Haule2009,deMedici2011,deMedici2011a}.
(ii) The orbitals appear to decouple from each other~\cite{deMedici2011a,deMedici2009,deMedici2014,Yin2011b,Fanfarillo2015} if one focuses on static correlators~\cite{Kugler2019}.
(iii) Spin dynamics appears to slow down at
low energies (``spin freezing'')~\cite{Werner2008}.
(iv) Various correlators show fractional power law 
behavior~\cite{Werner2008,Yin2012,Walter2020}.
(v) Correlations depend strongly on the value of $J$ and relatively
less strongly on the value of $U$. (vii) The interplay of 
spin and orbital degrees of freedom leads to ``spin-orbital separation'' (SOS) \cite{Yin2012,Aron2015,Stadler2015,Stadler2018,Deng2019,Kugler2019,Horvat2019,Walter2020,Wang2020}. Here, we focus particularly
on the latter phenomenon.

In an isolated atom, it is well known that $J$  simply aligns electronic spins in
different orbitals according to Hund's first rule
\cite{Hund1925}. But if the atom is hybridized with a metallic
environment, as in many multiorbital materials or impurity models, 
the effect of $J$ is much more intricate and subtle (and was, 
with a few exceptions \cite{Okada1973}, largely overlooked  or underestimated until this decade). Here, SOS emerges in a complex two-stage
Kondo-type screening process, in which spin screening occurs at much
lower energies than orbital screening \cite{Stadler2015,Stadler2018}:
$\Tspin < \Torb$ (cf.\ Appendix~\ref{appendix:asym} for 
precise definitions of these scales). The low-energy regime below $\Tspin$ is a Fermi liquid (FL) governed by Landau QPs with heavy masses. By
contrast, the intermediate energy window featuring SOS,
$[\Tspin,\Torb]$, is governed by almost fully screened orbital degrees
of freedom weakly coupled to almost free spin degrees of freedom,
leading to incoherent behavior. Its non-Fermi-liquid (NFL) properties
are caused by an underlying novel NFL fixed point, described in detail
in Refs.~\cite{Wang2020,Walter2020} for a 3-channel spin-orbital Kondo (3soK) model
for Hund metals, as suggested in \oRef{Aron2015}.

As a function of increasing temperature, SOS leads to a
coherence-incoherence crossover with a coherence scale that is
strongly suppressed by Hund's coupling \cite{Stadler2015}. 
The coherence-incoherence crossover was predicted
in material simulations of iron oxypnictides already in
2008 \cite{Haule2008, Haule2009}. It was observed
a few years later in measurements of the resistivity,
heat-capacity, thermal-expansion coefficients, susceptibility, and optical
conductivity of the 122-iron
pnictides \cite{Hardy2013,Hardy2016,Yang2017}.
Further, only recently \cite{Deng2019}, realistic material simulations
and model Hamiltonian studies of the temperature dependence of the
local spectrum and of the charge, spin, and orbital susceptibilities of
the Hund metal \sroone{} and the Mott material \vo{} revealed that, for
Hund metals, SOS also occurs in the onset (and completion) of
screening of the orbital and spin degrees of freedom: 
 as the temperature is lowered in Hund
metals, the static local orbital and spin susceptibilities show
deviations from Curie behavior at different scales: $\TspinO < \TorbO$.
By contrast, for  Mott materials we have $\TorbO \approx \TspinO$,
since both these scales are equal to the scale $\TM$ at which
the Mott gap closes when the temperature is lowered.

During the last years, many insights on SOS have been gained in the
context of a minimal 3-orbital Hubbard-Hund  model (\HHM) for Hund
metals. In Refs.~\cite{Stadler2015,Stadler2018,Walter2020} the focus
has mainly been on zero-temperature results, while some
finite-temperature results were published in \oRef{Deng2019}. In the
present paper, we build on and extend the latter study by providing a
full analysis of the \textit{temperature dependence} of \ARPESt{} spectra, 
spectral function, self-energy, static local spin and orbital
susceptibilities, the QP weight, scattering rate, resistivity,
thermopower, and entropy. We choose four different sets of 
system parameters, which
mimic the physics of a Hund system (\Hone), a Mott system (\Mone),
an intermediate system (\Itwo) showing aspect of both Hund and Mott physics,
and a weakly correlated system (\Wzero).  With this we aim to
clarify previously-proposed criteria and also identify new 
ones for distinguishing
the two distinct routes of screening from atomic degrees of
freedom towards emerging quasiparticles, guided by either Mott or Hund
physics.

This paper is structured as follows. First we introduce the \HHM in \Sec{sec:Mod}. In 
\Sec{sec:Set} we shortly review the current state of research on the \HHM and 
motivate our choice of model parameters. Sections~\ref{ARPES},~\ref{SUS}, and \ref{TRANS} present our results. 
Section \ref{ARPES} concentrates on  \ARPESt{} spectra, as well as spectral functions and self-energies. In particular, we discuss the 
different temperature dependencies of these quantities for Hund and Mott systems. Based on our discussion of the \ARPESt{} spectra, in \Sec{SUS}, 
we explain  in detail the behavior of the static local orbital and spin susceptibilities and the quasiparticle weight 
in terms of  the SOS screening process. In \Sec{TRANS} we analyze signatures of Hund and Mott systems in
various transport properties (scattering rate, coherence scale, resistivity, effective chemical potential,
thermopower). Further, we study the lattice entropy and 
demonstrate that it differs from the impurity entropy. 
Remarkably, we are able to calculate the lattice entropy directly from our numerical data.
We summarize our insights in \Sec{sec:Concl} by providing tables, which highlight the most important features for distinguishing Mott and Hund
physics. 
Appendix~\ref{sec:asymfdq} additionally offers a detailed analysis of the particle-hole
asymmetry of the \HHM at $T=0$ and of  the frequency and temperature dependence of 
the optical conductivity. Further, it contains elementary definitions of several quantities discussed in \Sec{TRANS}.

\section{Model and Method}
\label{sec:Mod}

The minimal \HHM model for Hund metals, first suggested in  \oRef{Yin2012}, is described by the Hamiltonian 
\begin{eqnarray}
&& \hat{H}  =  \sum_{i} \left(  - \mu \hat N_{i} 
+ \hat{H}_\interact [\hat d^\dag_{i\nu}] \right) 
+\sum_{\langle ij\rangle \nu} t\,
  \hat{d}^{\dagger}_{i\nu}\hat{d}^{\phantom{\dagger}}_{j\nu} ,
\label{eq:HU} \\
&& \hat{H}_\interact[\hat d^\dag_{i\nu} ] 
=  \tfrac{1}{2}\left( U-\tfrac{3}{2}J \right) 
\hat N_i (\hat N_i -1)-{J}\hat{\mathbf S}_i^2
+ \tfrac{3}{4} {J}  \hat N_i .
\text{}\notag 
\end{eqnarray}
The on-site interaction term incorporates Mott and Hund physics
through $U$ and $J$, respectively. $\hat d^\dagger_{i \nu}$ creates an
electron on site $i$ of flavor $\nu = (m\sigma)$, composed of
a spin ($\sigma \! = \uparrow,\downarrow$) and orbital ($m=1,2,3$)
index.  $\hat n_{i\nu} =\hat{d}^{\dagger}_{i\nu}\hat{d}^\pdag_{i\nu}$
counts the electrons of flavor $\nu$ on site $i$.
$\hat N_i =\sum_{\nu}\hat n_{i\nu}$ is the total number operator for
site $i$ and $\hat{\mathbf S}_i$ its total spin, with components
$\hat S_i^\alpha = \sum_{m\sigma\sigma'}\hat{d}^{\dagger}_{i m\sigma}
\tfrac{1}{2}\sigma^\alpha_{\sigma\sigma'}\hat{d}_{i m\sigma'}$,
where $\sigma^{\alpha}$ are Pauli matrices.  We take a uniform hopping
amplitude, $t=1$, serving as energy unit in the \HHM, and a Bethe
lattice in the limit of large lattice coordination. The total width of
each of the degenerate bands is $W=4$. 
We choose the chemical
potential $\mu$ such that the total filling per lattice site is
$n_d\equiv \langle N_i \rangle=2$, 
i.e., the  three degenerate bands host two electrons.
The effective bare gap of this model is given by 
$\Delta_b\equiv U-2J$. (For a motivation of this
definition, see \oRef{Stadler2018}.)
We emphasize that Hund's coupling
plays no role at filling $n_d=1$, unless
the Hund's coupling itself becomes so large
that it starts mixing orbitals with different
occupation. In the latter case, similar
Hund's signatures may be observed even for a
2-orbital model with possible relevance
to certain materials \cite{Ryee2021}.

We have solved the 3HHM of  \Eq{eq:HU} 
using dynamical mean-field theory (DMFT) \cite{Georges1996} combined with a
state-of-the-art multiband impurity solver, the full-density-matrix
numerical renormalization group
(fdmNRG) \cite{Weichselbaum2007,Weichselbaum2012b}, while fully exploiting
the model's U(1)${}_\charge\times$SU(2)${}_\spin\times$SU(3)${}_\orb$ symmetry
using the QSpace tensor library \cite{Weichselbaum2012a}.
This approach  has yielded valuable insights into the complex interplay of 
spin and orbital degrees of freedom before
\cite{Stadler2015,Stadler2018,Deng2019,Walter2020}, because it
delivers high-quality results directly on the real-frequency axes and
for all physically relevant energies and temperatures. Details of the
DMFT+fdmNRG method are described in
Refs.~\cite{Stadler2015,Stadler2018,Stadler2019}. Method-related parameters
are given in the Supplementary Material of \oRef{Stadler2015}.

\section{Background and Setup}
\label{sec:Set}

This paper is strongly based on the insights gained in
\oRef{Stadler2018} for the \HHM at $T=0$. In the following, we give a
short overview of the most important facts established there. These
will be used later to analyze the temperature dependence of various physical
quantities in the \HHM.
\paragraph*{Phase diagram.}
\begin{figure*}
\centering
\includegraphics[width=0.8\linewidth, trim=14mm 175mm 14mm 10mm, clip=true]{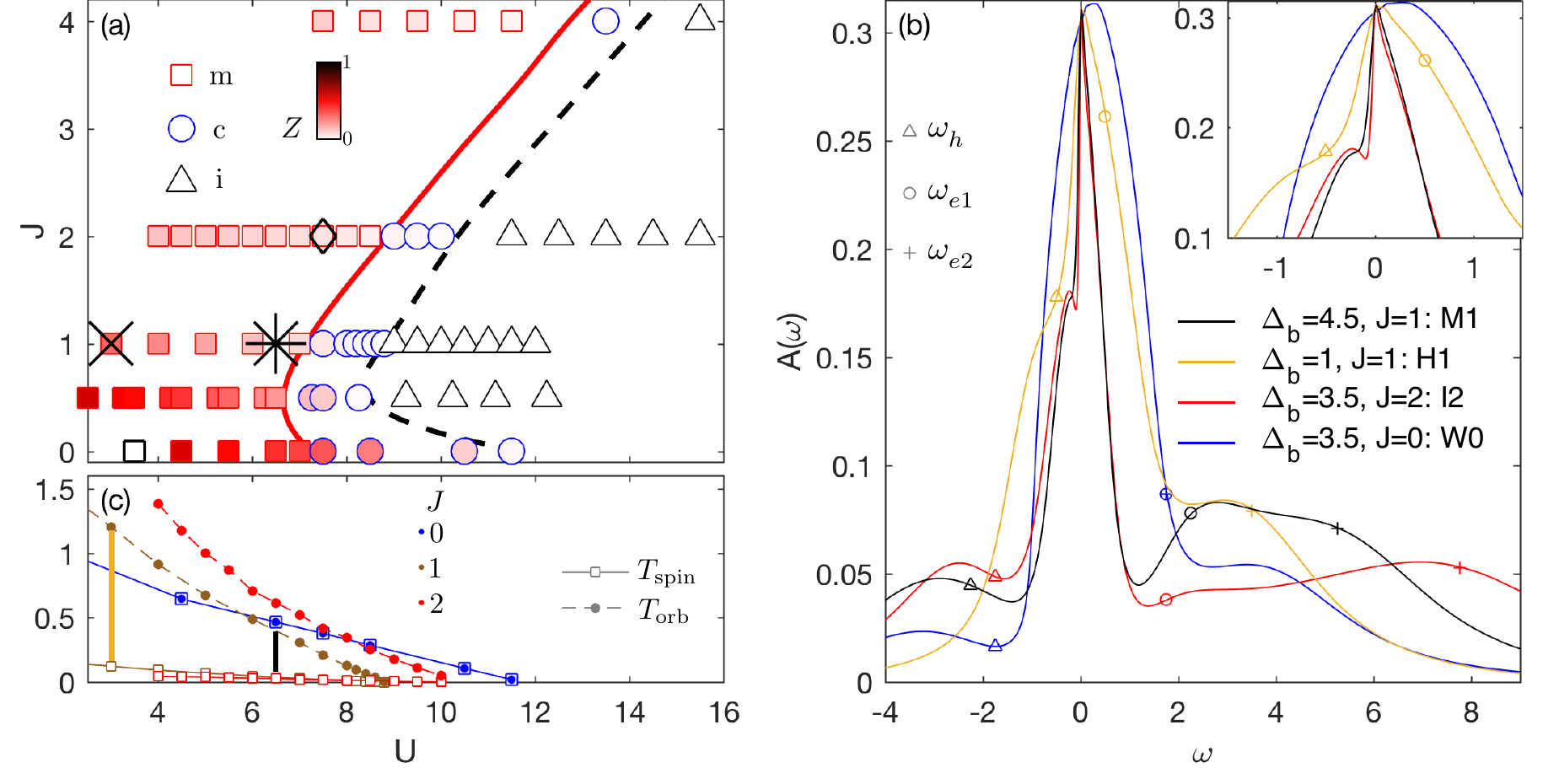} \vspace{-2mm}
\caption{(a) The zero-temperature phase diagram of the \HHM has three
     phases in the $J$-$U$ plane: a metallic phase (squares), a
     coexistence region (circles), and an insulating phase (triangles). 
     These are separated by two phase transition lines $U_{c1}$
     (solid red curve) and $U_{c2}$ (dashed black curve),
     respectively.  The color intensity of the symbols in the metallic and coexistence regions indicates the value of $Z\in[0, 1]$: the lower $Z$ the more faded is the red color. The phase diagram is adapted from \oRef{Stadler2018}. We will present temperature-dependent results for a Hund system \Hone{} far away from the $U_{c1}$ phase transition
     line deep in the metallic state (cross), a Mott system \Mone{} near the transition (asterisk), an intermediate system \Itwo{} having both Hund and Mott features (open diamond), and a  weakly correlated system \Wzero{} with $J=0$ far from $U_{c1}$ (open square).  (b) The local density of states $A(\omega)$ for  \Mone{} (black), \Hone{} (yellow), \Itwo{} (red), and \Wzero{} (blue). The legend
     lists the corresponding values of the bare gap, $\Delta_b = U - 2 J$.
     Triangles, circles, and crosses mark the bare atomic excitation scales, $\omega_h$, $\omega_{e1}$, and $\omega_{e2}$ (listed in increasing order), respectively, defined in Sec.~\ref{sec:Set}. 
     The inset zooms into the peaks around the Fermi level $\omega = 0$.
     (c) The spin and orbital Kondo scales, $\Tspin$ (solid) and $\Torb$ (dashed), plotted as function of $U$ for $J=0$ (blue),  $J=1$ (brown), and $J=2$ (red); these scales are defined
     as the maxima of the imaginary parts of the dynamic orbital and spin susceptibilities, see \App{sec:asymfdq}.
      The SOS window is marked by a vertical yellow (black) bar for \Hone{} (\Mone).}
\label{fig:phasediagram}
\end{figure*}
In \oRef{Stadler2018} we explored the \HHM at $1/3$ filling in a broad
region of parameters at $T=0$ and established the $J$-$U$ phase diagram,
replotted in \Fig{fig:phasediagram}(a).  It consists of three different
phases: a metallic phase (squares), a coexistence region (circles),
and an insulating phase (triangles), separated by two phase transition
lines $U_{c1}$ (solid red curve) and $U_{c2}$ (dashed black curve),
respectively. Thus, for fixed $J$, a Mott insulator transition (MIT)
occurs with increasing $U$, discussed extensively in
\oRef{Stadler2018}. The red color intensity of the symbols reflects the
strength of the quasiparticle weight, obtained from the
self-energy of the self-consistent lattice Green's function via
\begin{align}
Z= \frac{1}{{1-\left. \partial_{\omega} \real
    \Sigma(\omega)\right|_{\omega=0}}}= \frac{m}{m^\ast},
 \end{align}
with $m$ the free electron mass and $m^\ast$ the renormalized QP mass.
Importantly, for sizeable $J\gtrsim1$ (cf.\ \oRef{Stadler2018} for
details), strong correlation effects, i.e., considerable mass
enhancements $Z^{-1}$ occur not only close to the MIT lines but also
far from it (cf.\ e.g., faded red color for \Hone).

In \oRef{Stadler2018} we aimed to identify the origin of strong correlations far from and close to the MIT in \Fig{fig:phasediagram}(a).
To this end, we proposed several characteristic
signatures distinguishing Hund-correlated from Mott-correlated systems at
$T=0$. We briefly recapitulate the findings from \oRef{Stadler2018} in the following three paragraphs.

\paragraph*{Hund system.}   
The \HHM shows behavior typical of Hund metals at moderate and small $U$ values, i.e., far from a MIT phase boundary. 
As a prototypical example, we choose the Hund system \Hone{}
[marked by a cross in \Fig{fig:phasediagram}(a)] with $J=1$ and a small
bare gap $\Delta_b = 1$.  This choice relies on the fact that
\Hone{} qualitatively reproduces various physical properties of the
Hund metal \sroone{} \cite{Deng2019}. At $T=0$, Hund systems are characterized by the following signatures.

The lowest bare
atomic excitation scale
$E_{\rm atomic}=\omega_{e1}=-\omega_h=\tfrac{1}{2}U-J$ is typically
small due to the small value of $U$ and the sizable value of $J$
(e.g., $E_{\rm atomic}^{\Hone}=0.5$ for \Hone). The bare atomic scales,
$\omega_h$, $\omega_{e1}$, and $\omega_{e2}=\tfrac{1}{2}U+2J$ define
the characteristic energy scales, i.e., the peak positions, of
the Hubbard bands in the local density of states, 
\begin{align}
A(\omega)= -\tfrac{1}{\pi}\imag [G_\imp(\omega)],
\end{align}  
cf.\ yellow
crosses in \Fig{fig:phasediagram}(b). Thus, for \Hone, the Hubbard bands
form a broad incoherent background.
  
In Hund systems, strong correlations are induced by ``Hund physics'': The spin Kondo scale is strongly reduced due to SOS, with 
$\Tspin=0.12$ 
for \Hone{} [cf.\ brown curves in \Fig{fig:phasediagram}(c)]. Accordingly, the
QP mass, $Z^{-1}=3.45\propto{\Tspin}^{-1}$ \cite{Stadler2018}, 
is strongly enhanced. By contrast, $\Torb=1.20$
is even larger than
$E_{\rm atomic}=0.5$ for \Hone.  This leads to a very broad SOS frequency window $[\Tspin,\Torb]=1.08$
comparable in magnitude to $\Delta_b=1$ in Hund systems [cf.\ yellow vertical bar
in \Fig{fig:phasediagram}(c) for \Hone].  
The incoherent regime is
strongly particle-hole
asymmetric in frequency space 
\cite{Stadler2015,Stadler2018} and shows fractional power-law
behavior \cite{THLee2018,Wu2019,Stadler2019,Walter2020}.  At zero
temperature, the two-step SOS Kondo screening process is reflected in
$A(\omega)$ in form of a two-tier QP peak on top of the broad
incoherent background. It consists of a thin spin Kondo peak related
to spin screening and a broader orbital Kondo peak related to orbital
screening [cf.\ yellow curve in \Fig{fig:phasediagram}(b)] \cite{Stadler2018}.

\paragraph*{Mott system.}  
A Mott system is by definition close to the MIT phase boundary. $U$ is
large compared to $J$. We choose the Mott system \Mone{} [marked by an
asterisk in \Fig{fig:phasediagram}(a)] with $J=1$ and a large bare gap
$\Delta_b\equiv U-2J=4.5$ as a prototypical example. \Mone{}
qualitatively reproduces various physical properties of the
well-studied Mott system \vo{} \cite{Deng2019}.  The lowest bare atomic
excitation scales $\pm E_{\rm atomic}^{\Mone}=\pm 2.25$ are large
due to the large value of $U$, and the Hubbard bands therefore 
well
separated [cf.\ black curve in \Fig{fig:phasediagram}(b)]. By contrast, with
increasing $U$, both $\Torb$ and $\Tspin$ are linearly reduced, while
their ratio remains constant [cf.\ brown curves in
\Fig{fig:phasediagram}(c)]. As a consequence the SOS window
is strongly downscaled
$[\Tspin=0.04,\Torb=0.39]$,
becoming almost negligibly small compared to $\Delta_b=4.5$ 
[cf.\ black vertical bar in \Fig{fig:phasediagram}(c) for \Mone]. Since both Kondo scales are small, the QP peak is narrow altogether and well separated from the
Hubbard side bands [cf.\ black curve in \Fig{fig:phasediagram}(b)]. In sum,
Hund physics is only observable at very low energy scales. Typical Mott
physics, induced via the DMFT self-consistency, dominates.

\paragraph*{Absence of Hund's coupling.}   
For $J=0$, SOS is absent: spin and orbital degrees of freedom are
screened at the same scale, $\Tspin=\Torb$ [cf.\ blue curves in
\Fig{fig:phasediagram}(c)]. Far from the MIT phase boundary, e.g., for
\Wzero{} with $J=0$ and $\Delta_b=3.5$ [marked by an open square in
\Fig{fig:phasediagram}(a)], $\Tspin=\Torb= 0.7405$  are rather large
and thus $Z^{-1}= 1.5134$ not much enhanced: the system is only
weakly correlated. The QP peak has no substructure [cf.\ blue curve in
\Fig{fig:phasediagram}(b)].

\paragraph*{Temperature-dependence.} 
The size and the properties of the SOS window in frequency space has
direct implications for temperature dependent properties of the
\HHM. This was first demonstrated in \oRef{Deng2019}. In particular, it
was shown that, in local spectra, the QP peak persists up to very
high temperatures in Hund systems, exhibiting large charge fluctuations,
whereas a pseudogap develops with increasing temperature in all Mott
systems at a characteristic energy scale $\TM$, suppressing charge
fluctuations. This can be explained by the fact that far from the MIT boundary the Hubbard bands
overlap, whereas close to the boundary they are well separated. Furthermore, onset scales for orbital and spin screening,
  $\TorbO$ and $\TspinO$, were introduced as the scales where
  decreasing temperature first causes deviations of the respective
  static local orbital and spin susceptibilities, $\chiorb$ and
  $\chispin$, from the Curie behavior, $\chi \propto 1/T$,
  characterizing free local moments. In Hund metals, it was found
that $\TorbO\gg \TspinO$ with $\TorbO$ as high as $E_{\rm atomic}$. In
contrast, in Mott systems, spin and orbital screening set in, 
simultaneously,  below a much lower scale,
$\TspinO\approx\TorbO \approx \TM \ll E_{\rm atomic}$, together with
the formation of the QP peak.  A weakly correlated  system with
$J=0$ likewise does not exhibit any separation of the onset scales of
orbital and spin screening. 

In addition, completion scales for
orbital and spin screening, $\TorbC$ and $\TspinC$, were defined as
the temperature scale below which Pauli behavior sets in with
decreasing temperature. It was suggested that these scales are also
separated in the presence of finite $J$ in both Hund and Mott systems,
while they are equal for $J=0$ \cite{Deng2019}.

\paragraph*{Strategy.}
In the following, we analyze and compare four different systems, \Hone, \Mone, \Wzero{} and \Itwo, as presented in  \Fig{fig:phasediagram}(a), to
further clarify the Hund and Mott routes towards strong
correlations. The Hund system, \Hone, and the Mott system, \Mone, are defined as in \oRef{Deng2019}. In
addition, we also study the weakly correlated system \Wzero{}   
and  an intermediate system \Itwo{} with $J=2$ and
$\Delta_b=3.5$ [marked by an open diamond in
\Fig{fig:phasediagram}(a)], which has both Hund and Mott features and
thus demonstrates the crossover between Hund and Mott
systems.
For all these systems we summarize the physics in \ARPESt{} spectra at
$T=0$ and study their temperature dependencies. While some of this data is already presented as the Supplementary Information of \oRef{Deng2019}, we
here analyze it in much more detail and directly connect it to the
temperature dependence of various other physical quantities. In
particular, we revisit the static local susceptibilities and the idea
of completion and onset scales of spin and orbital
screening. Further insights are obtained  by studying the
quasiparticle weight, the resistivity, the thermopower, and the lattice
entropy. We will show that the latter differs from the impurity
entropy, studied before in \oRef{Stadler2015}.
In \App{sec:asymfdq}, we also offer a detailed
discussion, for \Itwo{}, of the implications of particle-hole
asymmetry for various frequency-dependent quantities at $T=0$.  All in all, these studies lead to a deepened
understanding of the nature of Hund metals.

\section{ARPES, spectral function, and self-energy}
\label{ARPES}
\begin{figure*}
\centering
\includegraphics[width=0.9\linewidth, trim=0mm 0mm 0mm 0mm, clip=true]{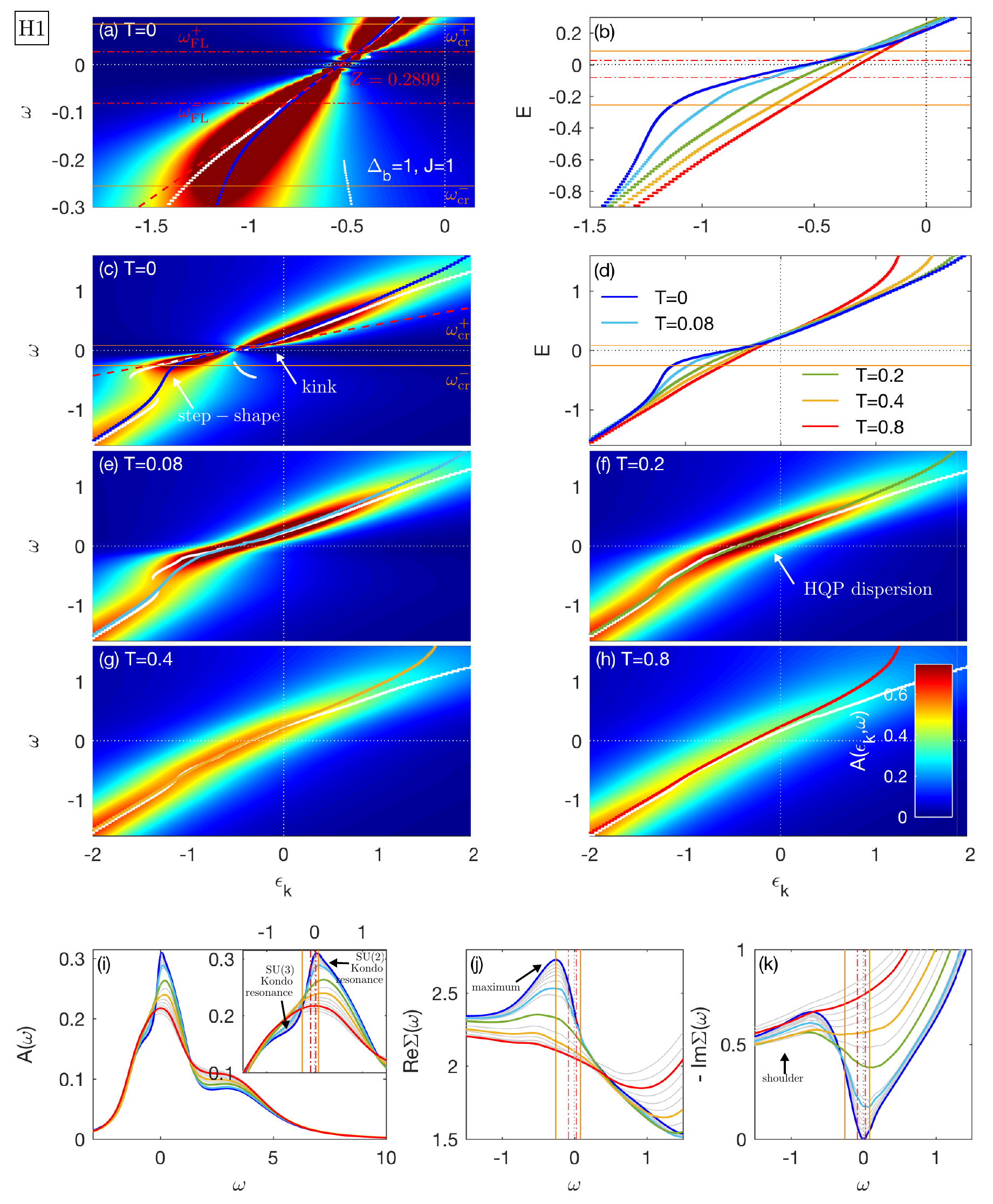}
\caption{ A Hund system (\Hone) with parameters $\Delta_b=1$
    and $J=1$.  [(a),(c),(e)--(h)] The structure factor $A(\epsilon_k,
    \omega)$. [(b),(d)] The dispersion relation $E(\epsilon_k)$, (i) the
    spectral function $A(\omega)$, [(j),(k)] the real and imaginary parts 
    of the self-energy, $\real\Sigma(\omega)$ and $\imag\Sigma(\omega)$, respectively, all plotted for various
    temperatures. [(a),(c),(e)--(h)]
    The colored curves highlight the dispersion relation
    $E(\epsilon_k)$ and the white curves show the alternative definition
    of the dispersion relation $E^*(\omega)$. Panels
    (a) and (b) are low-energy zooms of panels (c) and (d).  
   The FL regime, $\omega^{-}_\fl<\omega<\omega^{+}_\fl$, lies between the
 dash-dotted red lines, running horizontally in 
    (a) and (b) and vertically in (i)--(k). The thick dashed red
    line in panel (a) denotes FL behavior of the low-energy dispersion
    relation. Its slope $Z=m/m^\ast$ reflects the strength of local
    correlations. The yellow solid horizontal lines in (b) and 
 vertical lines in  (i)--(k) denote, for $\omega<0$, the energy scale
    $\omega^{-}_\cross$ of the maximum
     in $\real\Sigma_{T=0}(\omega<0)$, and for $\omega > 0$, 
   the energy scale $\omega^{+}_\cross$  of the kink
   in $\real\Sigma_{T=0}(\omega>0$).}
\label{fig:AkwDeltab1J1} 
\end{figure*}
In this section we focus on \ARPESt{} spectra.
We calculate the structure factor $A(\epsilon_k, \omega)$ for a Bethe
lattice as 
\begin{align}
\label{eq:ARPES}
A(\epsilon_k, \omega)= -\tfrac{1}{\pi}\imag\left[\omega + \mu -
  \epsilon_k-\Sigma(\omega)\right]^{-1} . 
\end{align}
Experimentally, the structure factor can be measured by angle-resolved photoemission spectroscopy (ARPES).
For brevity, our $A(\epsilon_k, \omega)$ spectra
will be called ARPES spectra, too, although they are of course
computed, not measured.
The four Figs.~\ref{fig:AkwDeltab1J1}, \ref{fig:AkwDeltab4_5J1},
\ref{fig:AkwDeltab3_5J2}, and \ref{fig:AkwDeltab3_5J0}
show our results for $A(\epsilon_k, \omega)$, together with the corresponding  
spectral function $A(\omega)$ and self-energy $\Sigma(\omega)$
for the four
systems \Hone, \Mone, \Itwo, and \Wzero, respectively.
$A(\epsilon_k, \omega)$ is plotted for
different temperatures in panels (a),(c), and (e)--(h). 
 $A(\omega)$ is plotted for several temperatures in panel (i), analogously, $\real\Sigma(\omega)$ in panel
(j), and $\imag\Sigma(\omega)$ in panel (k).  In the following, we are
particularly interested in how SOS is reflected in \ARPESt{} data at
$T=0$, and how it develops with increasing temperature in Hund
systems compared to Mott systems.  How can the emerging differences be
explained and interpreted physically?

\subsection{Hund system \Hone}

Let us first analyze \Fig{fig:AkwDeltab1J1}  for \Hone. Here, we start with the $T=0$ results [\Figs{fig:AkwDeltab1J1}(a) and \ref{fig:AkwDeltab1J1}(c)]. 
We reveal three  regimes with different behavior of the \ARPESt{} spectrum,
$A(\epsilon_k, \omega)$, due to SOS.

\paragraph*{Fermi-liquid regime at $T=0$.}
Figure \ref{fig:AkwDeltab1J1}(a) is a zoom into the FL regime, 
which at $T=0$ sets in for $|\omega|<\Tspin=0.1221$.  
The white curve shows the $\omega$ dispersion of the QP band,
defined as the maxima $E^{*}(\omega)$ of
$A(\epsilon_k, \omega)$ for given $\omega$, 
and the blue curve the $\epsilon_k$ dispersion,
defined as the maxima $E(\epsilon_k)$ of $A(\epsilon_k, \omega)$ for given $\epsilon_k$.
 Both definitions lead to the same low-energy linear
FL dispersion relation (cf.\ thick dashed red line) of slope $Z=0.29$
[with a Fermi surface crossing point $E^{*}(\omega=0)=\mu_\eff$].  The
mass enhancement of the Landau QPs in the Hund system \Hone{} is thus
fairly large, $Z^{-1}=m^\ast/m=3.45$. 
We define  $\omega^{-}_{\fl}$ and $\omega^{+}_{\fl}$
as the negative and positive crossover scales between which FL behavior holds [as diagnosed from a detailed
analysis of the $\omega$ dependence of $A(\omega)$ and $\Sigma(\omega)$, 
see \App{sec:asymfdq} for  a detailed discussion]. Interestingly, we find that the extent of the FL regime is different for negative or positive frequencies, $\omega^{-}_{\fl} \neq \omega^{+}_{\fl}$
(cf.\ thin dash-dotted red horizontal lines): the white (blue) QP band dispersion deviates earlier from the
thick dashed red FL line on the positive frequency side, i.e., at a
lower scale $\omega^{+}_\fl\approx-\tfrac{1}{3}\omega^{-}_{\rm
  FL}=0.027$. 
 The asymmetry of the FL regime directly
reflects the particle-hole asymmetry of the model away from
half-filling. The asymmetry of the FL regime
is discussed in more detail in Appendix~\ref{sec:asymfdq}. With
$\omega^{+}_{\fl}- \omega^{-}_{\fl}=0.109$, 
the FL
regime is rather large in \Hone{} (compared to the lowest bare atomic excitation scale $E_{\rm atomic}^{\Hone}=0.5$).  We remark that a similar
asymmetric FL regime was found earlier in a one-band hole-doped Mott
insulator~\cite{Deng2013}, i.e., for a particle-hole asymmetric model
with only one type of degrees of freedom (spins).  There, it was also
shown that a well-defined QP peak of ``resilient'' QP excitations
exists at temperatures above the FL scale, and that it dominates an
intermediate incoherent transport regime.  

\textit{Crossover regime at $T=0$.}  Above $\omega^{+}_{\fl}$ and below
$\omega^{-}_{\fl}$ the QP band starts to deviate from FL behavior and
crosses over into the NFL regime. In this regime, the dispersion
relation becomes highly particle-hole asymmetric, as clearly visible
in \Fig{fig:AkwDeltab1J1}(c). For $\omega>0$, $E$ (and $E^{*}$) turn
upwards with increasing $\epsilon_k$ into a steeper approximately
linear function. This crossover is reflected in a weak kink around a
crossover scale $\omega^{+}_\cross=0.085$ 
(solid yellow line at $\omega>0$).
For $\omega<0$, $E$ develops into a step-shaped curve for decreasing
$\epsilon_k$ approximately at the crossover 
scale $\omega^{-}_\cross=-0.256$
(solid yellow line at $\omega<0$). By contrast, $E^{*}$ essentially
keeps following the red FL line almost down to $\omega^{-}_\cross$,
before a jump signals the transition to a new type of 
transport regime, the HQP regime, where HQP stands for ``Hund quasiparticle", explained further below.

\textit{HQP regime at $T=0$.} 
For $\omega$ below the above-mentioned jump, i.e., well smaller
  than crossover scale 
  $-\omega^{-}_\cross$, the $\omega$ dispersion $E^{*}(\omega)$ 
  (white line) approaches the steep linear behavior of the 
  $\epsilon_k$ dispersion $E(\epsilon_k)$ (blue line).
Thus, the dispersion in the HQP regime is again linear, similar to the 
FL regime, but it is steeper than in the latter, 
for both negative and positive $\omega$.
This signals the survival of resilient but lighter QPs in the HQP regime, described in more detail below. 
Interestingly, the slope of $E$ ($E^{*}$) is slightly larger for
negative ($\omega< - \omega^{-}_\cross$) than for positive
($\omega>\omega^{+}_\cross$) frequencies, 
indicating different effective masses for electrons and holes.

\textit{SOS Kondo screening process.}
\begin{figure}
\centering
\includegraphics[width=1\linewidth, trim=0mm 0mm 0mm 0mm, clip=true]{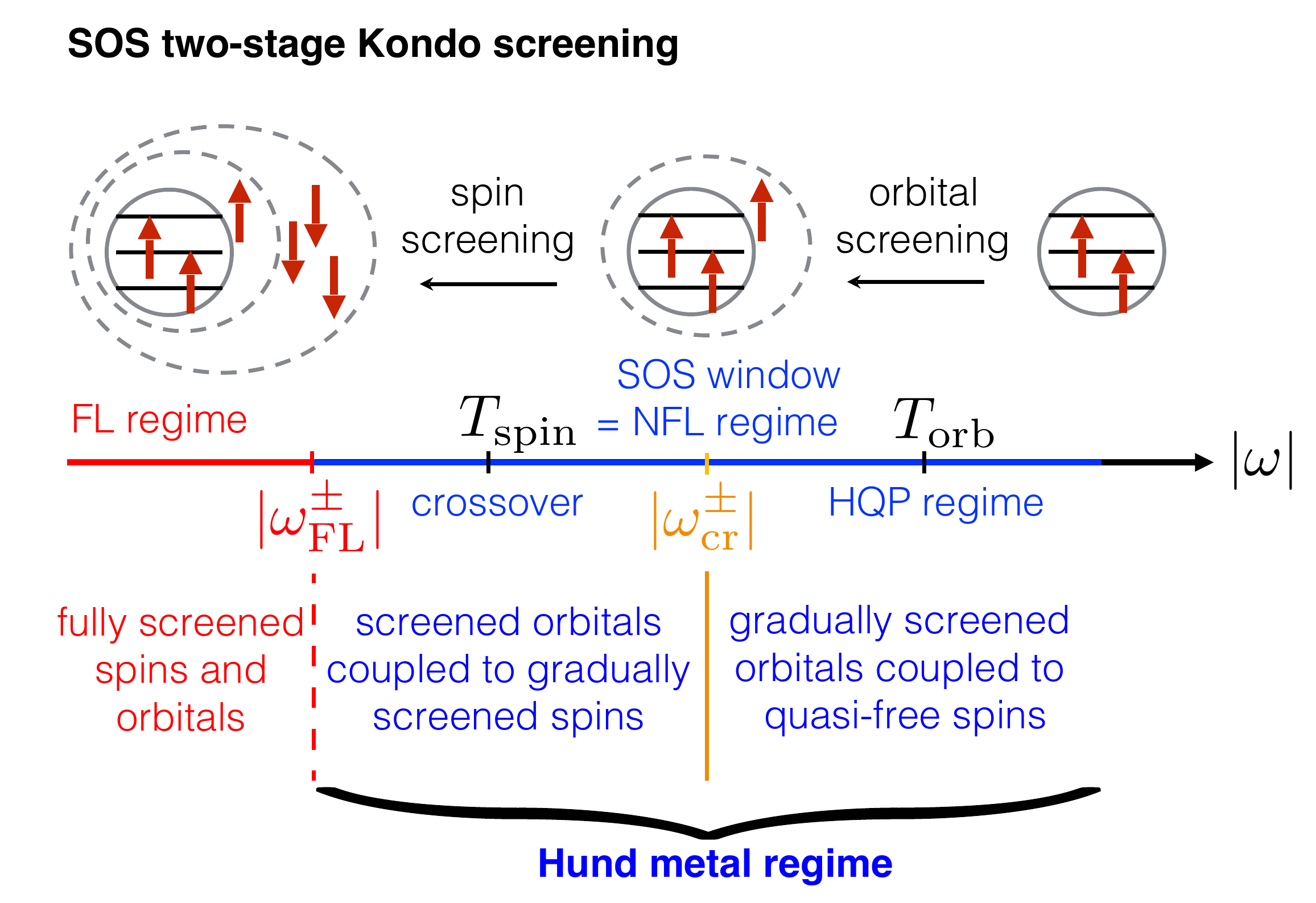}
\caption{Refined schematic depiction of the two-stage Kondo
  screening process of SOS at filling $n_d=2$ 
  (based on Fig.~13 of
  \oRef{Stadler2018}). With decreasing energy
  orbital screening sets in first, roughly 
   at the orbital Kondo scale $\Torb$. This involves the
  formation of an orbital singlet by building a large effective
  Hund's-coupling-induced ${3}/{2}$ spin including a bath spin degree
  of freedom. $|\omega^{\pm}_\cross|$ approximately marks the
  completion of orbital screening. 
  Below $|\omega^{\pm}_\cross|$ 
  the  ${3}/{2}$ spin is gradually screened by the three effective channels
  of the \HHM. Well below the spin Kondo scale $\Tspin$, full
  screening of both orbital and spin degrees of freedom is reached at
  the FL scale $|\omega^{\pm}_\fl|$, below which FL behavior
  occurs in frequency-dependent quantities. Our schematic sketch ignores the effects   of particle-hole asymmetry on the crossover scales, $|\omega^{-}_\cross|   \neq |\omega^{+}_\cross|$ and $|\omega^{-}_\fl|   \neq |\omega^{+}_\fl|$.}
\label{fig:scrsketch}
\end{figure}
We can now establish a connection between the three different
frequency regimes identified above in the \ARPESt{} spectrum, and the
intertwined two-stage Kondo screening process of SOS
(cf.\ \Fig{fig:scrsketch}) analyzed in
Refs.~\cite{Stadler2015,Stadler2018,Deng2019}.  Proceeding from high to low
frequencies (energies), orbital screening sets in first. This 
involves the formation of an orbital singlet, by binding one bath
electron to the impurity to screen the orbital hole. Due to
Hund's coupling, the extra bath electron couples ferromagnetically to the impurity, leading to the emergence of a large effective ${3}/{2}$ impurity spin. This transport regime has NFL properties,
but is characterized by an \ARPESt{} spectrum with a surprisingly linear
band dispersion, having a much steeper slope, i.e., a much smaller
mass enhancement, than in the FL regime. It might thus be described in
terms of specific resilient QPs, which are formed by gradually
screened orbital degrees of freedom coupled to quasi-free large
spins. We dub these resilient QPs ``Hund quasiparticles'' (HQPs).  The
steep slope of this HQP band (especially at negative frequencies) is
reminiscent of the (inverted) waterfall structure discovered in ARPES
spectra and realistic density functional theory (DFT) plus quantum Monte Carlo (QMC) studies of \sroone~\cite{Stricker2014}. We thus corroborate the suggestion of \oRef{Stricker2014} that the waterfall structure is a signature of resilient QPs in Hund metals.  But we also remark that a waterfall structure
was also found in ARPES plots for the hole-doped one-band Hubbard
model in \oRef{Deng2013}.  The ``completion'' of the orbital screening
process is reflected in a (strong) change in the band dispersion
around $\omega^{-}_\cross$ (step-shape) and $\omega^{+}_\cross$
(kink), respectively. Notably, subtle changes (kinks) at about 30 meV
were reported in ARPES data of
\sroone~\cite{Mravlje2011,Stricker2014,Tamai2019}, presumably caused by local electronic correlations~\cite{Tamai2019}, and
therefore could be associated with the crossover from the NFL to the
FL regime. For frequencies
below $\omega^{+}_\cross$ and above $\omega^{-}_\cross$ spin
screening sets in: the large ${3}/{2}$ spin is now screened by the
three channels of the \HHM to additionally form a spin singlet in the
ground state. Figuratively speaking the HQPs get additionally dressed
by the spin degrees of freedom. After completion, FL behavior
characterizes the low-frequency regime. Here, the QP band can be
described in terms of Landau QPs with a heavy mass $Z^{-1}=m^\ast/m$,
reflected by the small slope $Z$ of the band dispersion in
\ARPESt{} data.  These Landau QPs are more stable on the negative
frequency side.  

As has been discussed in \Sec{sec:Set} and
Refs.~\cite{Stadler2015,Stadler2018}, the two-step screening process of SOS
is also reflected in $A(\omega)$ and $\Sigma(\omega)$. In $A(\omega)$
a narrow SU(2) spin Kondo peak sits on top of a broad SU(3) orbital
Kondo peak [cf.\ blue curve in \Fig{fig:AkwDeltab1J1}(i)], resulting in a
shoulder for $\omega< \omega^{-}_\cross$ and a subtle kink for
$\omega>\omega^{+}_\cross$ (cf.\ vertical solid yellow lines).
Correspondingly, $-\imag\Sigma(\omega)$ [cf.\ blue curve in
  \Fig{fig:AkwDeltab1J1}(k)] develops a shoulder below
$\omega^{-}_\cross$ and a regime above $\omega^{+}_\cross$ in
which the slope of $-\imag\Sigma(\omega)$ becomes smaller than for
$\omega<\omega^{+}_\cross$. 
The scattering rate in the HQP regime is thus less energy dependent 
than in the FL regime.  The
shoulder-like structure in $-\imag\Sigma(\omega<0)$ directly
translates to a sharp maximum in $\real\Sigma(\omega<0)$ [cf.\ blue curve
  in \Fig{fig:AkwDeltab1J1}(j)]. We use the position of this maximum
to define $\omega^{-}_\cross$ (vertical solid yellow line at
$\omega<0$). The kink in $\real\Sigma(\omega>0)$ approximately
  marks $\omega^{+}_\cross$ (vertical solid yellow line at $\omega>0$),
  which turns out to lie at $\tfrac{1}{3}\omega^{-}_\cross$. 
  While these scales are in
principle heuristic choices, their physical relevance can be motivated
by the fact that they directly reflect the energy scales of marked
changes in the band dispersion $E(\epsilon_k)$: the latter is the
solution to the equation $\omega+\mu-\epsilon_k-\real
\Sigma(\omega)=0$, as used in \oRef{Deng2014}, and thus directly
connected to $\real \Sigma(\omega)$.  In Appendix~\ref{sec:asymfdq},
we complement this discussion by a detailed investigation of the
frequency dependence of $A(\omega)$, $\Sigma(\omega)$, and the
dynamical spin and orbital susceptibilities, $\chi_{\spin}(\omega)$
and $\chi_{\orb}(\omega)$, at $T=0$ for the system \Itwo{} and their
interpretation in terms of the SOS screening process. 

We remark that the SOS features described above, in particular
the shoulder below $\omega_\cross^-$ in both $A(\omega) $ and $-\imag\Sigma(\omega)$, have also been predicted
to occur for {Sr}${}_2${MoO}${}_4$ in very recent DFT+DMRG studies
\cite{Karp2020}.
\paragraph*{Temperature dependence.}
 In order to verify the idea of robust HQPs governing the 
 incoherent transport regime, we study the evolution of the QP band
 and its dispersion $E$ with temperature in
 \Figs{fig:AkwDeltab1J1}(c), \ref{fig:AkwDeltab1J1}(e)--\ref{fig:AkwDeltab1J1}(h) and \Figs{fig:AkwDeltab1J1}(b) and \ref{fig:AkwDeltab1J1}(d),
 respectively.  We find that, with increasing temperature, first the
 SOS features in the dispersion, like the step at
   $\omega<0$ and the kink at $\omega>0$, dissolve gradually and very
   slowly, while the steep slope of the linear behavior characteristic
   of the HQP regime 
   remains unchanged
 [cf.\ \Fig{fig:AkwDeltab1J1}(d)].
At $T\gtrsim0.2$ 
 the Landau-FL QP band has fully disappeared and only a slight kink at
 the Fermi level separates the linear parts of the resilient HQP band
 at $\omega>0$ and $\omega<0$ [cf.\ green curves in
   \Figs{fig:AkwDeltab1J1}(b), \ref{fig:AkwDeltab1J1}(d), and \ref{fig:AkwDeltab1J1}(f)]. The slope
 of the HQP band remains quite stable over a very broad range of
 frequencies (especially for $\omega<0$) up to the highest temperature
 plotted [cf.\ \Fig{fig:AkwDeltab1J1}(d)]. 
 Thus the incoherent transport regime for $T\gtrsim0.2$ is 
 governed by a very robust, almost temperature independent HQP band.  

 This evolution of the QP band with increasing 
 temperature is also reflected in
 $A(\omega)$, $\real\Sigma(\omega)$, and $\imag\Sigma(\omega)$ [cf.\
   \Figs{fig:AkwDeltab1J1}(i)--\ref{fig:AkwDeltab1J1}(k)]. In the FL temperature regime a sharp
 $\textrm{SU(2)}$ Kondo peak in $A(\omega)$, a pronounced maximum in
 $\real\Sigma(\omega)$, and a shoulder and dip in
 $\imag\Sigma(\omega)$ are clearly visible (cf.\ blue curves).  With
 increasing temperature there is a gradual crossover to NFL behavior. 
 The height of the $\textrm{SU(2)}$ Kondo resonance in
 $A(\omega)$ decreases 
and the two-tier structure of the QP peak
 disperses by redistributing spectral weight from the $\textrm{SU(2)}$
 Kondo peak to the $\textrm{SU(3)}$ Kondo resonance shoulder.
 However, the width of the broad $\textrm{SU(3)}$ Kondo resonance is
 essentially unaffected by this redistribution.  In fact, the
 robustness of the HQP band is reflected in the stable form of the QP
 peak flank of $A(\omega)$, especially at negative frequencies
 [cf.\ \Fig{fig:AkwDeltab1J1}(i)]. Interestingly, this flank is
 stabilized by the lower Hubbard band, which lies around
 $\omega_h=-0.5$, i.e., the $\textrm{SU(3)}$ Kondo resonance and atomic
 excitations merge in \Hone, resulting in a robust \ARPESt{}
 spectrum with mixed valence character at very high temperatures \cite{Deng2019}.
 
Next we consider the self-energy. Reflecting
the temperature dependence of
$A(\omega)$, also the maximum in $\real\Sigma(\omega)$ and the dip and
the shoulder in $\imag\Sigma(\omega)$ get first gradually smeared out
with increasing temperature for $T\lesssim0.2$. Notably, the minimum of $-\imag\Sigma(\omega,T)$ is shifted
to positive frequencies within this process. This hints towards
long-lived electron-like excitations governing the incoherent
transport of this crossover regime. The minimum in
$-\imag\Sigma(\omega,T)$ disappears 
at higher temperatures and $-\imag\Sigma(\omega,T)$ becomes a
monotonically increasing function of frequency close to the Fermi level. This might
again be caused by mixed valence physics, which becomes important at
an energy scale of around $0.5$. 

Interestingly, very similar behavior of
the minimum of $-\imag\Sigma(\omega,T)$ is observed for the hole-doped
one-band Hubbard model of \oRef{Deng2013}. There, a well-defined QP
peak persists with increasing temperature above the coherence scale
until it merges with the lower Hubbard band at high temperatures.

Note that the temperature dependence of $\real\Sigma(\omega)$ directly
determines the temperature dependence of the dispersion relation
$E(\epsilon_k)$ in $A(\epsilon_k, \omega)$
[cf.\ \Fig{fig:AkwDeltab1J1}(d)].
 Again, the evolution of the QP band with temperature strongly hints towards the existence of different types of QPs. At very low $T$ in the FL regime, the band is described by a low-frequency FL-like QP band with a rather flat dispersion. Correspondingly, $A(\omega)$ exhibits a  sharp $\textrm{SU(2)}$ Kondo resonance. Then, with increasing temperature, a crossover takes place: The low-frequency FL-like QP band dissolves gradually until, at higher temperatures, we find a new QP regime, the HQP regime. There, a much steeper (slightly particle-hole asymmetric) HQP band exists and the two-tier QP peak in $A(\omega)$ is reduced to a  single broad resilient $\textrm{SU(3)}$ Kondo resonance.

 \subsection{Mott system \Mone}
 We now turn to the Mott system \Mone.
\begin{figure*}
\centering
\includegraphics[width=0.9\linewidth, trim=0mm 0mm 0mm 0mm, clip=true]{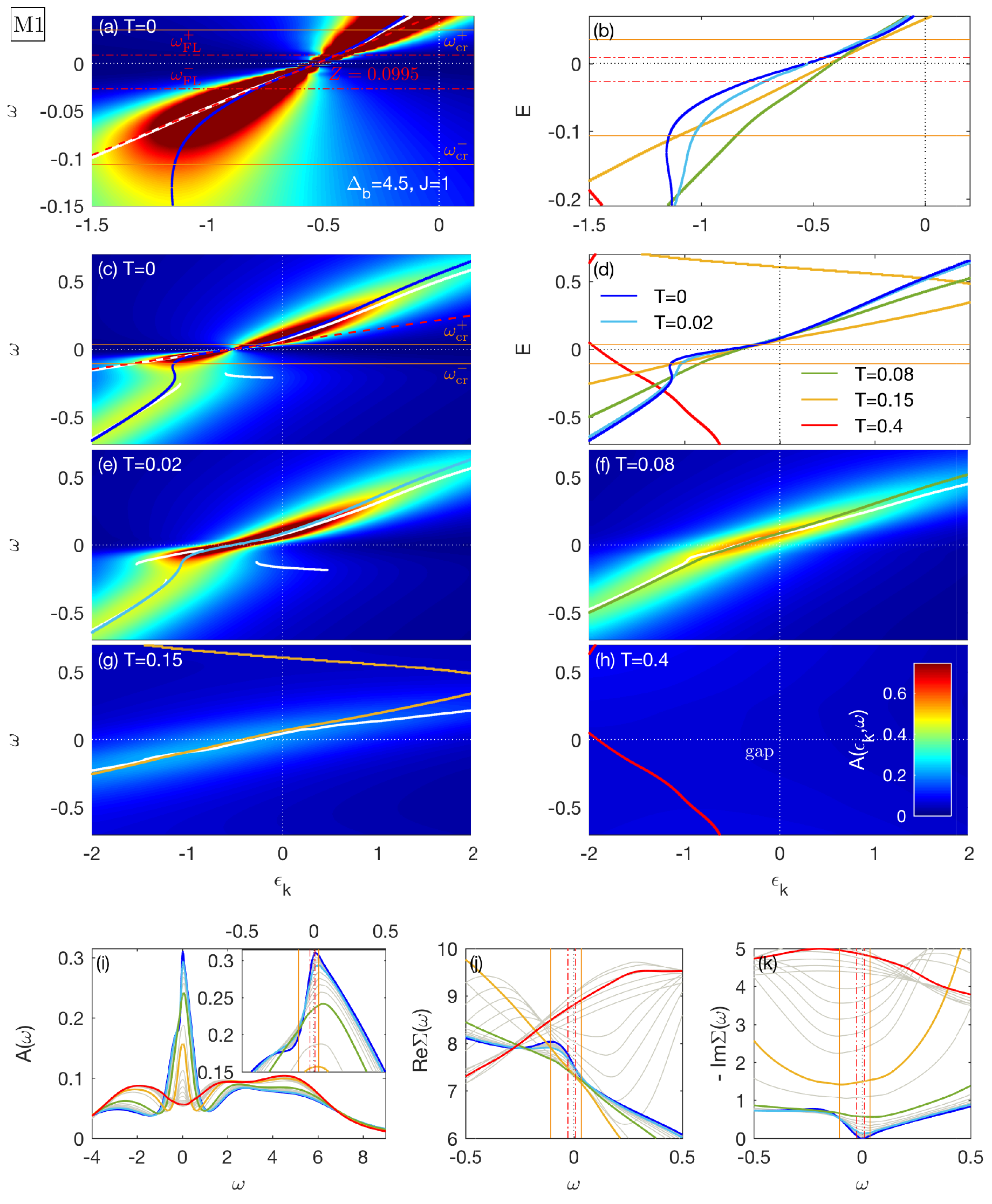}
\caption{Same quantities as in \Fig{fig:AkwDeltab1J1} for a Mott system (\Mone) with parameters $\Delta_b=4.5$ and $J=1$.}
\label{fig:AkwDeltab4_5J1}
\end{figure*} 
Figure \ref{fig:AkwDeltab4_5J1} displays its spectral properties using
the same layout as \Fig{fig:AkwDeltab1J1} for \Hone. At $T=0$ we again find a particle-hole asymmetric  FL frequency
regime and SOS features 
[cf.\ \Figs{fig:AkwDeltab4_5J1}(a) and \ref{fig:AkwDeltab4_5J1}(c)]. However, these occur at much
lower frequencies than in \Hone{} (for instance, M1 has $\omega^{-}_{\rm
  cr}=-0.15$), 
  as expected from the
insights given in \Sec{sec:Set}. The slope of the FL dispersion $Z=0.10$
is clearly smaller for \Mone{} than for \Hone{}, indicating much heavier electron masses. With increasing temperature, the SOS features vanish very quickly (already below $T=0.08$ for \Mone) [cf.\ \Figs{fig:AkwDeltab4_5J1}(b) and \ref{fig:AkwDeltab4_5J1}(d)]. The emergent HQP band [cf.\ \Fig{fig:AkwDeltab4_5J1}(f)] is very unstable with increasing temperature and already starts to disappear at around $T=0.15$ [cf.\ \Figs{fig:AkwDeltab4_5J1}(d) and \ref{fig:AkwDeltab4_5J1}(g)]. Above $T\gtrsim0.2$ a pseudogap
has fully replaced the QP peak [cf.\ \Figs{fig:AkwDeltab4_5J1}(h)].
Similarly, the whole QP peak in $A(\omega)$ becomes strongly 
suppressed, 
eventually turning into a pseudogap at high temperatures [red curve in
  \Fig{fig:AkwDeltab4_5J1}(i)]. The emergence of a pseudogap is accompanied by a change of sign, from positive to negative, in the slope of the dispersion relation $E(\epsilon_k)$ [cf.~red curve in Fig.~\ref{fig:AkwDeltab4_5J1}(d)].
  Consequently, $\real\Sigma(\omega)$
and $\imag\Sigma(\omega)$ are strongly temperature dependent, as
well. While for $T\lesssim0.08$ the minimum of
$-\imag\Sigma(\omega,T)$ is shifted to positive frequencies, it is
gradually shifted back towards negative frequencies with increasing
temperature and finally turns over to a maximum in the presence of a
pseudogap [cf.\ \Fig{fig:AkwDeltab4_5J1}(k)].

\subsection{Intermediate system \Itwo}
 \begin{figure*}
\centering
\includegraphics[width=0.9\linewidth, trim=0mm 0mm 0mm 0mm, clip=true]{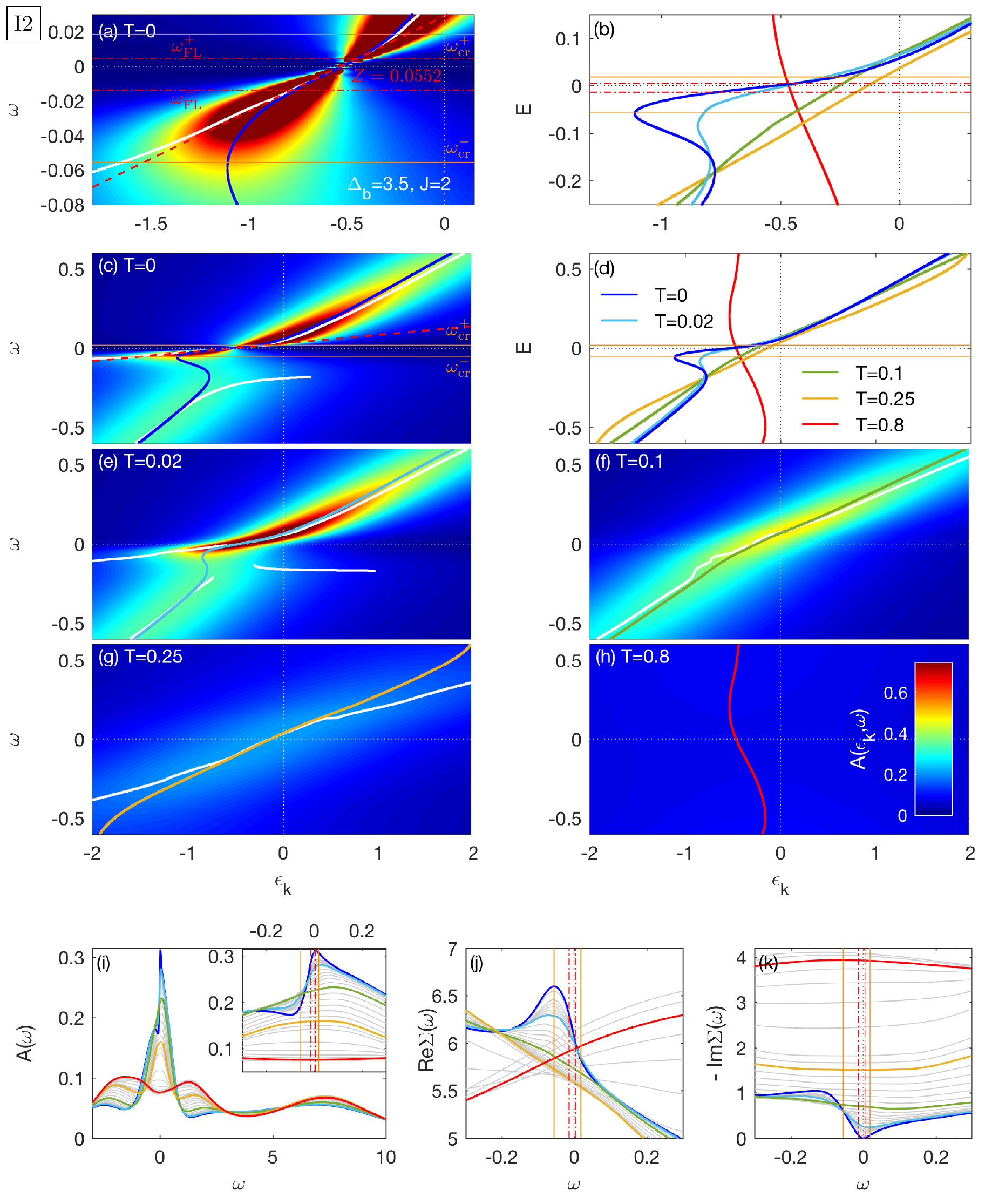}
\caption{Same quantities as in \Fig{fig:AkwDeltab1J1} for an intermediate system (\Itwo) with paramters $\Delta_b=3.5$ and $J=2$. }
\label{fig:AkwDeltab3_5J2}
\end{figure*}
Figure \ref{fig:AkwDeltab3_5J2} shows spectral data
for the intermediate system \Itwo{}. At $T=0$, 
the \ARPESt{} spectrum for \Itwo{} [cf.\ \Figs{fig:AkwDeltab3_5J2}(a) and \ref{fig:AkwDeltab3_5J2}(c)] shows SOS features similar to those of \Hone, but occurring at smaller scales. 
Since $J=2$ and the bare gap $\Delta_{b}=3.5$ are both large,  $\Tspin$ is pushed down \cite{Stadler2018} even compared to \Mone: 
$\Tspin=0.021$ and thus $Z=0.055$
 (cf.\ thick dashed red line) but also $|\omega^{\pm}_\fl|$ and $|\omega^{\pm}_\cross|$ take approximately half the values of the respective scales of \Mone, while $\Torb=0.42$
for \Itwo{} is slightly larger than $\Torb=0.3878$ for \Mone. 
In sum, the zero-temperature band dispersion of \Itwo{} is similar in its shape to \Hone{} and \Mone. 

However, qualitative differences  emerge in the temperature evolution of the QP band and its dispersion $E$ compared to \Hone{} and \Mone, respectively---again due to the specific relation  $[\Tspin,\Torb]/\Delta_{b}$ for \Itwo.
With increasing temperature, first the band's step-shaped structure gradually dissolves, while its steep linear behavior in the 
HQP frequency regime remains unchanged [cf.
  bright blue curve for $T=0.02$ in \Figs{fig:AkwDeltab3_5J2}(b), \ref{fig:AkwDeltab3_5J2}(d), and
  \Fig{fig:AkwDeltab3_5J2}(e)]. In contrast to \Mone{}, this HQP band is stable up to rather high temperatures, $T=0.25$, for \Itwo{} (similar to \Hone). Nevertheless, above $T=0.25$, we additionally find a crossover to a pseudogap similar to  \Mone{} [cf.\  red curves in
  \Figs{fig:AkwDeltab3_5J2}(b), \ref{fig:AkwDeltab3_5J2}(d), \ref{fig:AkwDeltab3_5J2}(h), and \ref{fig:AkwDeltab3_5J2}(i)].
\Itwo{} is thus characterized by both a Hund feature (HQP band) at
intermediate temperatures and a Mott feature (pseudogap) at very high
temperatures.
This evolution of the QP band with temperature is again reflected in
$A(\omega)$, $\real\Sigma(\omega)$, and $\imag\Sigma(\omega)$ [cf.
  \Figs{fig:AkwDeltab3_5J2}(i)--\ref{fig:AkwDeltab3_5J2}(k)].

\subsection{Weakly correlated system \Wzero} 
 \begin{figure*}
\centering
\includegraphics[width=0.9\linewidth, trim=0mm 0mm 0mm 0mm, clip=true]{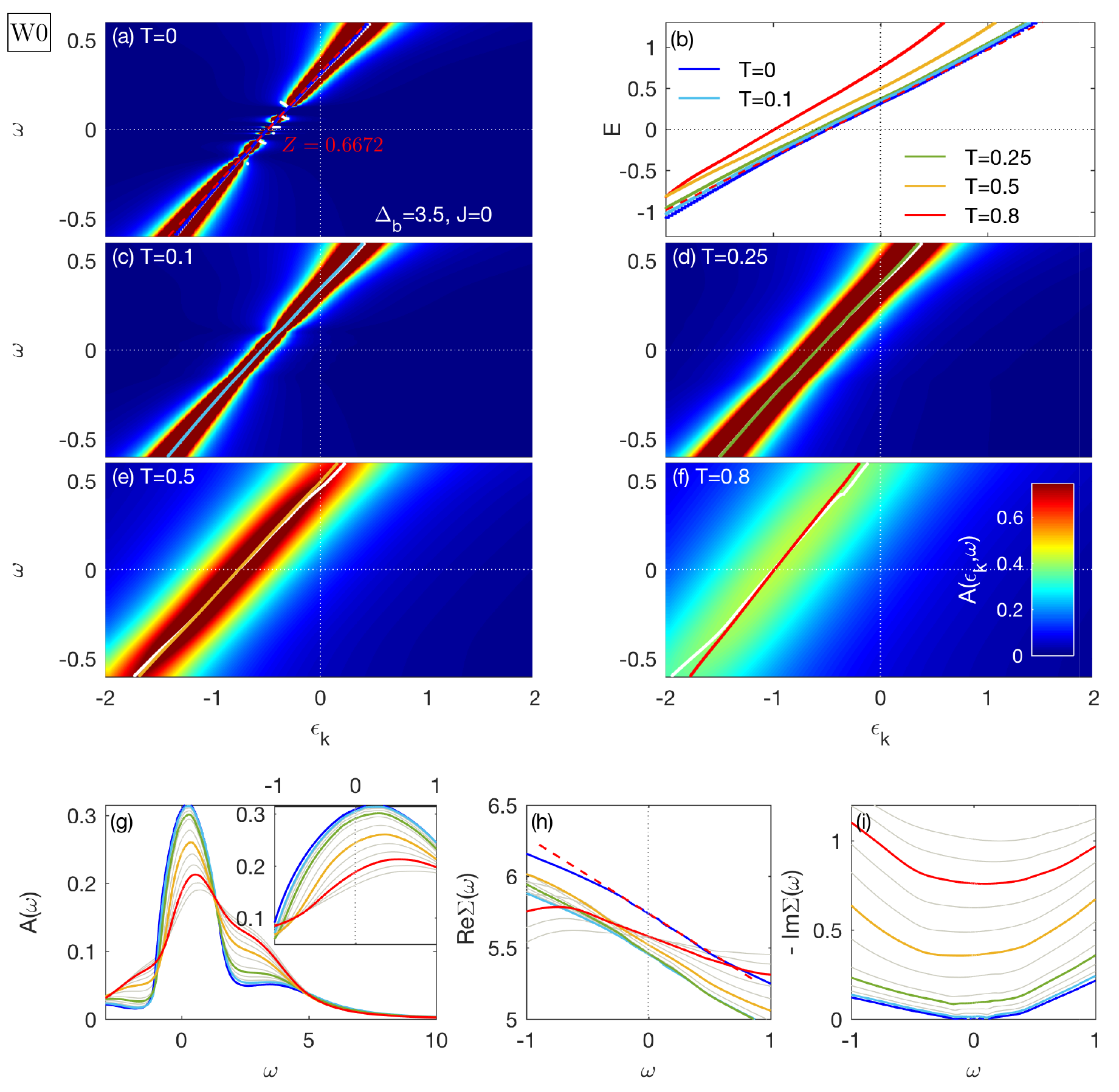}
\caption{A weakly coupled system \Wzero{} with parameters $\Delta_b=3.5$ and $J=0$. [(a),(c)--(f)] The structure factor $A(\epsilon_k,
    \omega)$. (b) The dispersion relation $E(\epsilon_k)$, (g) the
    spectral function $A(\omega)$, [(h),(i)] the real and imaginary parts 
    of the self-energy, $\real\Sigma(\omega)$ and $\imag\Sigma(\omega)$, respectively, all plotted for various
    temperatures. (h) Note
  that the difference in $\real\Sigma(\omega=0)$ between $T=0$ and
  $T>0$ arises from a $4\%$ deviation of $n_d(T=0)$ from $n_d=2$. FL and crossover scales are not shown. Note that the latter do not
  exist for $J=0$.}
\label{fig:AkwDeltab3_5J0}
\end{figure*}
For $J=0$ the SOS features are fully absent in $A(\epsilon_k, \omega)$, $A(\omega)$, $\real\Sigma(\omega)$, and $\imag\Sigma(\omega)$ (cf.\ \Fig{fig:AkwDeltab3_5J0}).
 The FL behavior holds for a rather large temperature regime (almost up to $T\approx0.25$)  
 and is characterized by  a very stable large dispersion with $Z=0.6672$  and thus a rather small mass enhancement. Resilient HQPs do not exist.

\subsection{Summary of spectral properties}
To summarize,  both \Hone\ and \Mone\ (and also \Itwo) show SOS features  in the dispersion extracted from $A(\epsilon_k, \omega)$ at $T=0$: (i) a rather flat low-frequency Landau QP band of slope $Z$; (ii) a NFL crossover behavior (in form of a step-shaped band  at $\omega<0$ and a kink at $\omega>0$); and (iii) a HQP band, which is extended in frequency space. The latter consists of positive and negative frequency parts, both of which exhibit linear dispersion relations with large slopes, with
the negative-frequency slope  slightly larger than the 
positive-frequency slope. However, these SOS features occur at very different energy scales for the three systems \cite{Stadler2018}: while in  \Hone\ they are extended over a broad frequency range up to atomic energy scales, they are compressed and lie at smaller frequency scales in  \Mone. Consequently, in  \Hone, these features govern 
transport for all temperatures.  In particular, very robust HQPs
exist up to the highest $(\lesssim t)$  temperatures. By contrast,
in \Mone, SOS physics only survives at very low temperatures, whereas
the behavior of $A(\epsilon_k, \omega)$ at higher temperatures is
dominated by typical Mott physics, i.e., the DMFT self-consistency
opens a (pseudo)gap and quickly destroys the HQPs.  For \Itwo, the SOS
features are also found at rather low scales (due to the large $\Delta_{b}$) at $T=0$, but the SOS regime is more extended than for \Mone{} (due to the large $J$). 
Temperature-dependent \ARPESt{} spectra thus show both Hund and Mott features.  If
$J=0$, SOS features are absent and \Wzero{} is governed by FL behavior
in a broad temperature range.

\section{Static local orbital and spin susceptibilities, and quasiparticle weight}
\label{SUS}
\begin{figure*}
\centering
\includegraphics[width=0.8\linewidth, trim=5mm 85mm 0mm 0mm, clip=true]{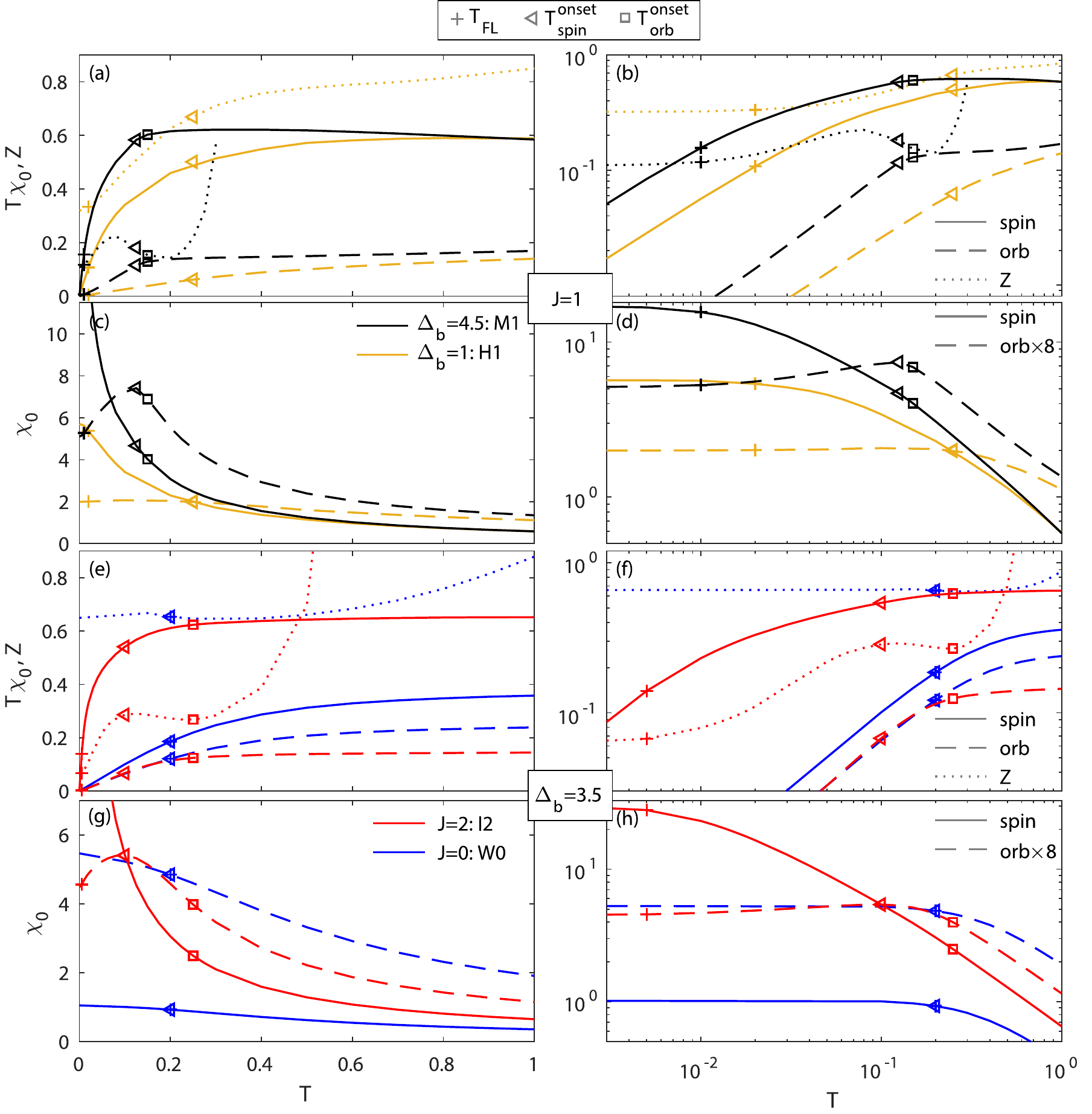}
\caption{The static local orbital (dashed) and spin (solid)
  susceptibilities, [(a),(b),(e),(f)] $T\chi^{\orb, \spin}$ and [(c),(d),(g),(h)]
  $\chi_0^{\orb, \spin}$, all plotted
  as functions of temperature on  a linear (left) and 
a logarithmic (right) scale, for \Mone{}
  (black), \Hone{} (yellow), \Itwo{} (red), and \Wzero{} (blue). In
  addition, the quasiparticle weight (dotted) $Z(T)$ is shown in
  (a), (b), (e), and (f).  The squares mark the onset of orbital screening
  $\TorbO$ below which $T\chi_0^{\orb}$ deviates from a constant
  value, i.e., from Curie-like behavior. Note that $Z(T)$ diverges for
  $T>\TorbO$. The triangles mark the maxima of $\chi_0^{\orb}$ and
  also signal the onset of spin screening $\TspinO$ below which
  $T\chi_0^{\spin}$ deviates from Curie-like behavior. The crosses
  denote the FL scale $T_\fl$ below which FL behavior is found. In
  \Mone, we observe that $\TspinO\approx \TorbO=\TM$. In \Hone{}, we
  find $\TorbO\gg\TspinO$, as discussed in \oRef{Deng2019}. The data for the static susceptibilities
shown in panels (a)--(d) are adapted from \oRef{Deng2019}. A Curie-Weiss
analysis of the data of panel (c) is presented in \App{app:CrossoverScales}.}
\label{fig:Tdep_1}
\end{figure*}

Based on the above detailed analyzis of the \ARPESt{} spectra, we now revisit the static local  susceptibilities for the orbital and spin degrees of freedom, to refine the findings which we had reported in \oRef{Deng2019}.
There we introduced four temperature scales, characterizing the onset and the completion of screening of the spin and the orbital degrees of freedom. 
The concept of onset and completion scales for screening
was inspired by Wilson's classic analysis of the impurity contribution to the
spin susceptibility of the spin-1/2 one-channel Kondo model, reviewed in \App{app:CrossoverScales}.
We correspondingly derived these scales from the behavior of the static local spin and orbital susceptibilities, and also of the 
local spectral function.  Our main result was that Hund and Mott systems show contrasting behavior at intermediate to high
energies. In Hund systems, we found a clear separation in the energy scales at which the screening for orbital and spin fluctuations sets in, respectively: $\TorbO \gg \TspinO$,  with $\TorbO$ very large 
($\gtrsim E_{\rm atomic}$).
By contrast, in Mott systems the strong Coulomb repulsion localizes the charge at high temperature. With decreasing temperature the onset of charge
localization triggers the simultaneous onset of the screening of the spin and orbital degrees of freedom, accompanied by the
formation of the coherence resonance at $\TM\equiv\TspinO=\TorbO\ll E_{\rm atomic}$. At low temperatures, we suspected
SOS in the completion of screening, $\TorbC\gg \TspinC$, both for Hund and Mott systems, but considered this to be more
pronounced for Hund systems.

In this section we now reanalyze the static local  susceptibilities of \Hone{} and \Mone{} of \oRef{Deng2019}. While we only slightly refine the onset scales of screening quantitatively to provide a clearer connection to corresponding \ARPESt{} data and the quasiparticle weight, we suggest a revised perspective on the completion scales. In sum, we establish a consistent physical picture of screening from the atomic degrees of freedom at high energies to the quasiparticles at low energies. We corroborate our findings by studying the static local  susceptibilities of \Itwo{} and \Wzero.

The dynamical real-frequency spin
and orbital susceptibilities  are defined as
\begin{subequations}
\label{subeq:susceptibility}
\begin{align}
\chi^{\spin}(\omega) & = \tfrac{1}{3}\sum_\alpha \langle \hat
S^\alpha \mbox{$\parallel$} \hat S^\alpha \rangle_\omega, 
\\ 
\chi^{\orb}(\omega) & = \tfrac{1}{8} \sum_a \langle \hat
T^a\mbox{$\parallel$} \hat T^a \rangle_\omega, 
\end{align}
\end{subequations}
respectively
\cite{Hanl2014,Weichselbaum2012b}, where
$\hat T^a = \sum_{mm'\sigma}\hat{d}^{\dagger}_{m\sigma}
\tfrac{1}{2}\tau^a_{mm'}\hat{d}_{m'\sigma}$
are the impurity orbital operators with the SU(3) Gell-Mann matrices
$\tau^a$ normalized as ${\rm Tr}[ \tau^a\tau^b ] =2\delta_{ab}$.
Below the subscript $0$ will be used to denote the static limit,
$\chi_0 = \chi(\omega=0)$, i.e., the static local  susceptibilities.

We plot $T\chi_0^{\orb, \spin}$ in \Figs{fig:Tdep_1}(a), \ref{fig:Tdep_1}(b), \ref{fig:Tdep_1}(e), and \ref{fig:Tdep_1}(f) and
$\chi_0^{\orb, \spin}$ in \Figs{fig:Tdep_1}(c), \ref{fig:Tdep_1}(d), \ref{fig:Tdep_1}(g), and \ref{fig:Tdep_1}(h) as functions of 
$T$, for \Hone{}
(yellow), \Mone{} (black), \Itwo{} (red), and \Wzero{} (blue).
As a function of decreasing temperatures, 
these susceptibilities traverse four regimes:
first Curie-like behavior, where $T\chi_0$ is independent of temperature;
onset of screening, where $T\chi_0 $ begins to decrease;
completion of screening, where $\chi_0$ begins to 
saturate; and Pauli behavior, where $\chi_0$ is
constant. We will discuss these regimes in detail below.

 We also plot the quasiparticle weight $Z(T)$ as dotted lines in
\Figs{fig:Tdep_1}(a), \ref{fig:Tdep_1}(b), \ref{fig:Tdep_1}(e), and \ref{fig:Tdep_1}(f) [and additionally in
  \Figs{fig:Tdep_2}(a) and \ref{fig:Tdep_2}(b)]. In principle,  
the interpretation of $Z(T)$ as quasiparticle weight holds only
  in the FL regime.  Nevertheless, for temperatures in the NFL
regime, it is still computationally well-defined and we use it to
interpret the physics on a heuristic level.

\subsection{Hund system \Hone}
\label{sec:SusceptibilityH1}

We begin with a discussion of the results for \Hone{} in
\Figs{fig:Tdep_1}(a)--\ref{fig:Tdep_1}(d). 
$T\chi_0^{\orb}$ decreases with decreasing
temperature for all temperatures plotted [cf.\ dashed yellow curves in
\Figs{fig:Tdep_1}(a) and \ref{fig:Tdep_1}(b)], i.e., the onset for orbital screening,
$\TorbO>1$, is on the order of bare excitation scales.  The onset of
spin screening, $\TspinO\approx0.25$, is signaled by the deviation
from Curie-like (constant) behavior of $T\chi_0^{\spin}$ with
decreasing temperature, marked by the yellow triangle
[cf.\ solid yellow curves in \Figs{fig:Tdep_1}(a) and \ref{fig:Tdep_1}b)]. 
Thus, for \Hone, we find $\TorbO\gg \TspinO$, as shown in \oRef{Deng2019}. 
Note, however, that here we have chosen $\TspinO\approx0.25$ slightly smaller than in \oRef{Deng2019} (where we had chosen $\TspinO\approx0.4$). This choice is
motivated by the \ARPESt{} data in \Fig{fig:AkwDeltab1J1}. There the onset
of spin screening is reflected in the formation of a flat
low-frequency band in addition to the steep HQP band, resulting in a
pronounced step-like feature in the dispersion at $T=0$. In \Fig{fig:AkwDeltab1J1}(d) the onset of the step formation is visible for $T\lesssim0.2$.
Furthermore, we motivate our choice in terms of the behavior of
$\chi_0^{\orb}$. With $\TspinO\approx0.25$, the onset scale of spin
screening is equal to the temperature scale for the completion of
orbital screening: $\chi_0^{\orb}$ shows Pauli (constant) behavior for
$T<\TspinO\approx T^{\cmp}_{\orb}$ [cf.\ dashed yellow curve in
\Figs{fig:Tdep_1}(c) and \ref{fig:Tdep_1}(d)]. When the temperature is further lowered, 
$\chi_0^{\spin}$ too reaches Pauli behavior at $T_{\fl}=T^{\cmp}_{\spin}$
(yellow cross).  Then spin screening is completed and the system is a
FL.
\begin{figure}
\centering
\includegraphics[width=1.0\linewidth, trim=0mm 0mm 0mm 0mm, clip=true]{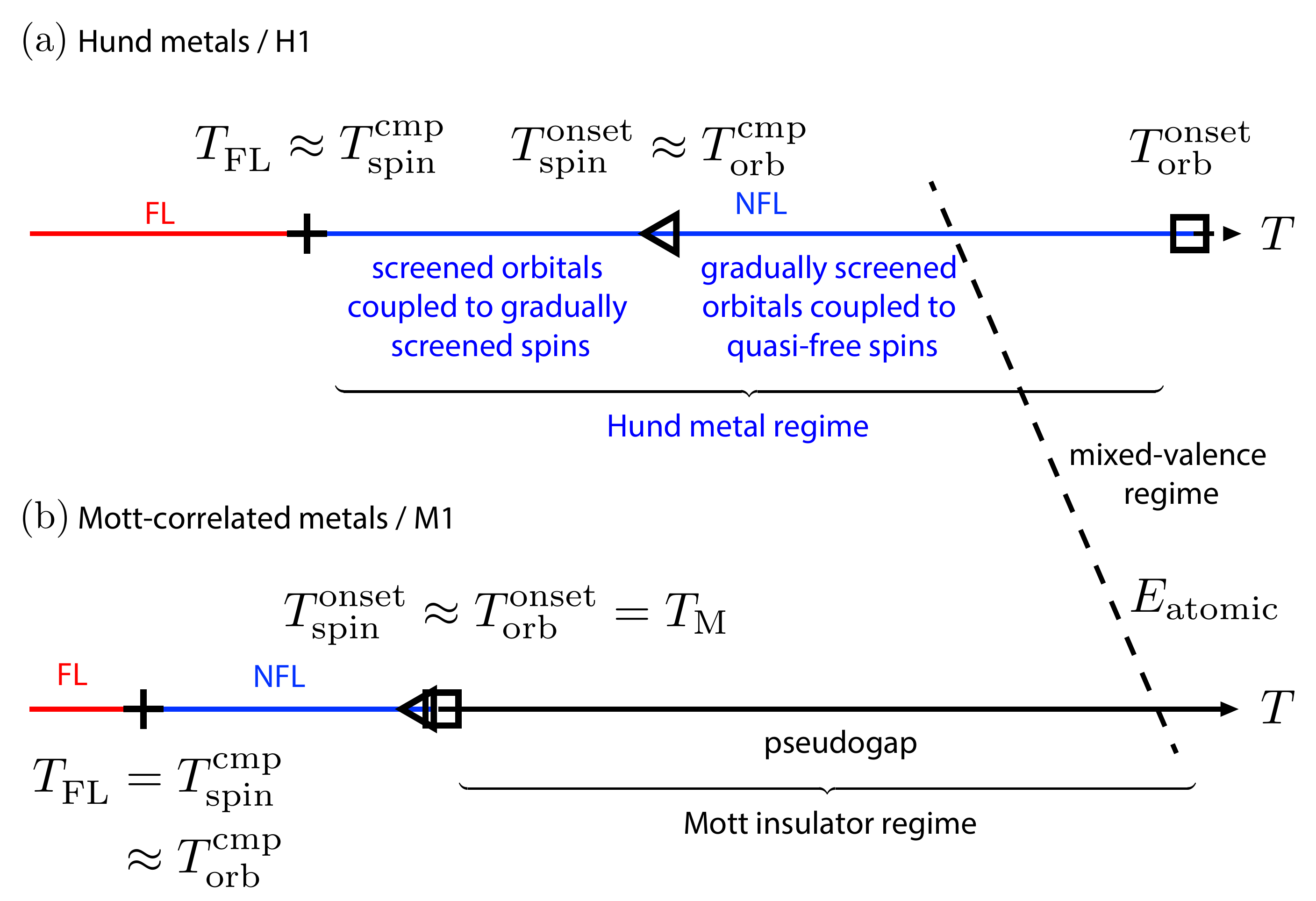}
\caption{(a) Schematic sketch of different temperature regimes in
a Hund metal.  For $T>T_\fl$, \Hone{} is a NFL up to temperatures in
the order of bare energy scales, where also mixed-valence physics
becomes important. The NFL  regime, which we dub Hund metal regime, reflects the complex SOS screening
process of \Fig{fig:scrsketch}. First orbitals get screened with
decreasing temperature for 
$\TorbC < T < \TorbO$. In this regime
transport is governed by HQPs, which are characterized by gradually
screened orbitals coupled to quasi-free spins. Only when the orbital
screening process is completed spins get screened below
$\TspinO\approx T^{\cmp}_{\orb}$, i.e., in this regime, the HQPs get
gradually dressed to form heavier Landau QPs.  For
$T<T_\fl=\TspinC$, \Hone\ is a FL and both orbital and spin
degrees of freedom are fully screened.  (b) Schematic sketch of
different temperature regimes in a multiorbital Mott-correlated
metal. In a Mott system, a pseudogap governs the physics in an extended
temperature regime, $T>\TspinO\approx \TorbO=T_{\textrm{M}}$. For
temperatures below $T_{\textrm{M}}$, both orbital and spin degrees of
freedom get screened simultaneously with the onset of a Kondo
resonance, which is driven by the DMFT self-consistency condition. The
NFL regime for $T_\fl < T < \TM$
is followed by a low-temperature FL regime, 
$T < T_\fl < \TspinC \approx \TorbC$, 
where
both orbital and spin degrees of freedom are fully screened.}
\label{fig:scrsketchT}
\end{figure}

Figure \ref{fig:scrsketchT}(a) summarizes these observations in a schematic
sketch. In a Hund system, the SOS screening process of
\Fig{fig:scrsketch} is directly reflected in the temperature
dependence of the static local susceptibilities. For
$\TspinO < T < \TorbO$, HQPs, i.e., gradually screened (quasi-itinerant)
orbitals coupled to quasi-free spins, dominate the physics and lead to
a robust HQP band in \ARPESt{} spectra and a Curie-like spin
susceptibility.  At very high temperatures mixed-valence physics 
additionally comes into play \cite{Deng2019}, because the lower (and a part of the
upper) Hubbard band merge at $\omega_h=-0.5$ (and
$\omega_{e1}=+0.5$) into the QP peak in \Hone{}
[cf.\ \Fig{fig:AkwDeltab1J1}(i)].  Due to the special SOS screening
process, the spin screening only sets in once orbital screening
has been completed $T^{\cmp}_{\orb}$, thus $\TspinO\approx
T^{\cmp}_{\orb}$. As the temperature is
lowered into the regime $T_{\fl} < T < \TspinO$ also the spins get gradually
screened, eventually resulting in the full screening of 
both spin and orbital degrees
of freedom and thus in a FL below $T_{\fl}$. The spin screening is
signaled by the formation of a step-like feature in \ARPESt{} spectra and
by a Pauli-like orbital susceptibility.

This screening route is also reflected in $Z(T)$ [cf.\ dotted yellow
  curve in \Figs{fig:Tdep_1}(a) and \ref{fig:Tdep_1}(b)]. For $\TspinO < T < \TorbO$, the
existence of resilient HQPs leads to a plateau-like feature in $Z(T)$.
As the temperature decreases into the regime 
$T_{\fl} < T < \TspinO$, $Z(T)$ decreases and
approaches a second plateau in the FL regime $T<T_{\fl}$. The
reduction of $Z(T)$ shows that the HQPs are additionally ``dressed''
through spin screening, resulting in heavier Landau QPs.

\subsection{Mott system \Mone}

The Mott system \Mone{} behaves very differently. As shown in
\oRef{Deng2019}, $T_{\textrm{M}}\equiv \TorbO\approx \TspinO$
    [cf.\ black triangle and square in \Figs{fig:Tdep_1}(a) and \ref{fig:Tdep_1}(b)]. For
    $T>T_{\textrm{M}}\approx0.15$, both $T\chi_0^{\orb}$ and
    $T\chi_0^{\spin}$ exhibit a Curie plateau and the spectral
    function is characterized by a pseudogap. Both spin and
      orbital degrees of freedom get screened simultaneously with the
    onset of a Kondo resonance [cf.\ \Figs{fig:AkwDeltab4_5J1}(g) and \ref{fig:AkwDeltab4_5J1}(i)],
    which is driven by the DMFT self-consistency condition, in
    contrast to the Kondo screening in \Hone{}. Interestingly,
    $\TspinO$ now corresponds to the position of a maximum in
    $\chi_0^{\orb}$ [cf.\ black triangle and black dashed curve in
      \Figs{fig:Tdep_1}(c) and \ref{fig:Tdep_1}(d)]: the orbital dynamics is strongly
    influenced by the spin screening and true Pauli behavior is only
    reached for $T<T_\fl$ in \Mone{}, thus
    $T_\fl=T^{\cmp}_{\spin}\approx T^{\cmp}_{\orb}$ for Mott systems.

    The behavor described above 
is summarized in \Fig{fig:scrsketchT}(b). In
    \Mone{}, Mott physics dominates and with increasing temperature
    essentially destroys SOS physics by opening a pseudogap already at
    low temperatures.  Again, $Z(T)$ reflects these findings
    [cf.\ dotted black curve in \Figs{fig:Tdep_1}(a) and \ref{fig:Tdep_1}(b)]. Similar to
    \Hone{}, $Z(T)$ is small and constant for $T<T_\fl$. But instead
    of a second HQP plateau as in \Hone{}, $Z(T)$ has a maximum
    directly below $T_{\textrm{M}}$ and diverges for
    $T>T_{\textrm{M}}$.

\subsection{Intermediate system \Itwo}

To corroborate our picture above, we similarly study \Itwo{} and
\Wzero{} in \Figs{fig:Tdep_1}(e)--\ref{fig:Tdep_1}(h). \Itwo{} is rather close to the Mott
boundary [cf.\ diamond in \Fig{fig:phasediagram}(a)]. Thus, we observe
Mott signatures at high temperatures: for $T>T_{\rm M}\equiv\TorbO\approx0.25$,
$T\chi_0^{\orb}$ shows Curie behavior [cf.\ red square in
\Figs{fig:Tdep_1}(e) and \ref{fig:Tdep_1}(f)] and a pseudogap exists [cf.
\Figs{fig:AkwDeltab3_5J2}(g) and \ref{fig:AkwDeltab3_5J2}(i)]. However, due to the large $J=2$, we
find Hund signatures, as well, at intermediate and low temperatures:
orbital and spin screening are slightly separated, $\TorbO> \TspinO$,
and $Z(T)$ features a plateau for $\TspinO < T <  \TorbO$ (between red
triangle and square). $\TspinO$ marks
a maximum in $\chi_0^{\orb}$ [cf.\ \Figs{fig:Tdep_1}(g) and \ref{fig:Tdep_1}(h)], which is
however less pronounced than for \Mone{}. Full screening with Pauli
behavior of both $\chi_0^{\orb}$ and $\chi_0^{\spin}$ is reached at
$T<T_\fl$. Due to the large Hund's coupling, $T_\fl$ (and accordingly
$Z(T=0)$~\cite{Stadler2018}) is lowest in \Itwo\ compared to \Hone,
\Mone, and \Wzero.   In sum, \Itwo{} exhibits an intermediate system, showing
a mixture of Hund and Mott features. 

\subsection{Weakly correlated system \Wzero}

Finally, we consider the weakly correlated system \Wzero{}, a system without Hund's coupling,
$J=0$ (cf.\ also \Fig{fig:AkwDeltab3_5J0}). In \Figs{fig:Tdep_1}(e)--\ref{fig:Tdep_1}(h),
$\chi_0^{\orb}$ and $\chi_0^{\spin}$ behave similarly 
for \Wzero, up to a constant prefactor:
$\chi_0^{\spin}/\chi_0^{\orb}=1.5$. The FL regime extends up to
very high temperatures [$Z(T)$ is essentially constant in an extended regime in \Figs{fig:Tdep_1}(e) and \ref{fig:Tdep_1}(f)]. Both Hund and Mott features are absent in
\Wzero{}.

\section{Transport properties and entropy}
\label{TRANS}

In this section we add further perspective to the 
differences and similarities of the four systems
\Hone, \Mone, \Itwo, and \Wzero\ by discussing the 
temperature dependence of various 
transport properties and the entropy. For completeness,
Appendix~\ref{app:definitions}
 collects some elementary definitions and relations
involving the quantities discussed below.

\begin{figure*}
\centering
\includegraphics[width=0.8\linewidth, trim=14mm 50mm 14mm 0mm, clip=true]{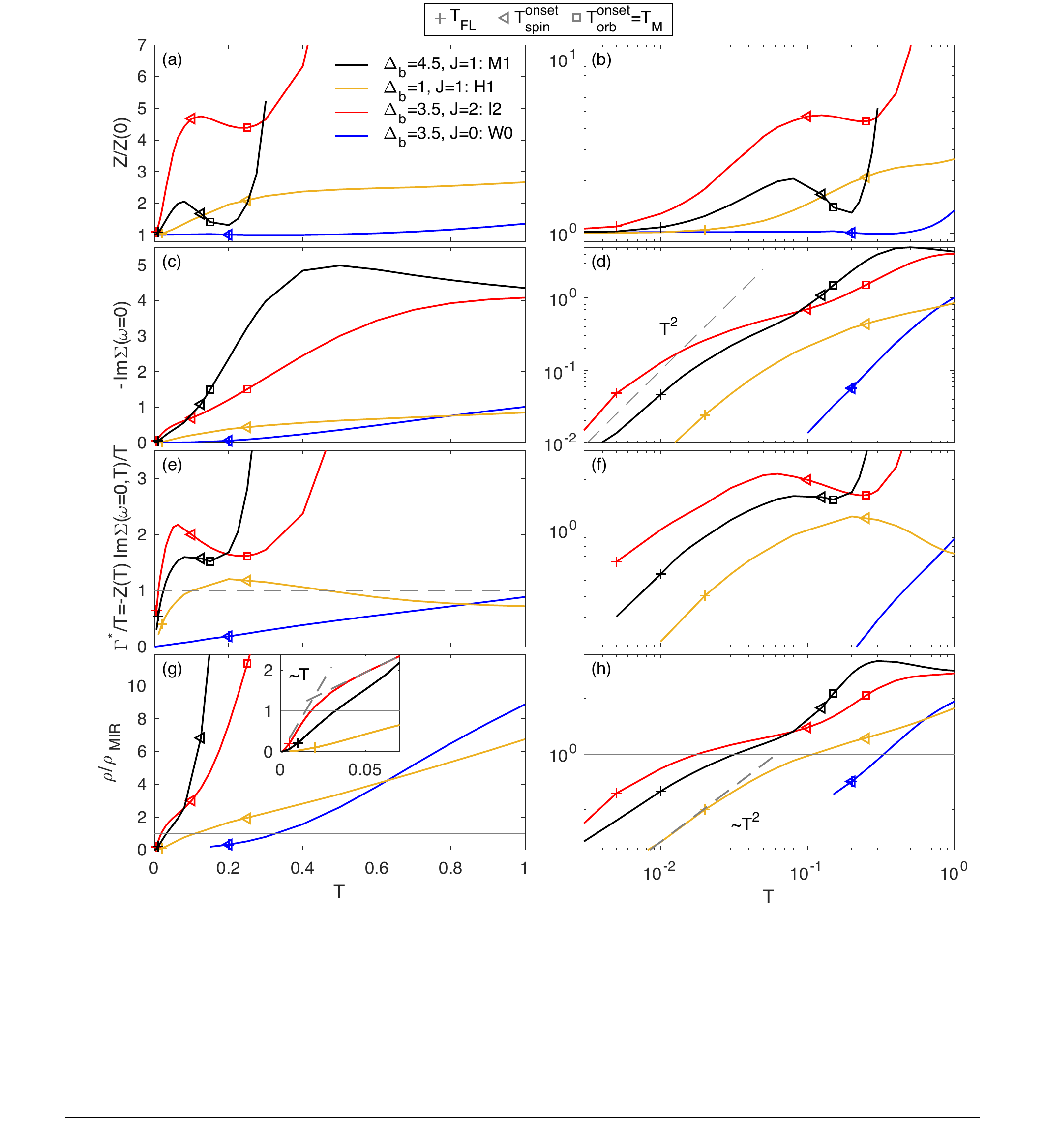}
\caption{[(a),(b)] The quasiparticle weight $Z/Z(0)$ (replotted from
    Fig.~\ref{fig:Tdep_1} for reference), [(c),(d)] the scattering rate
  at the Fermi level $\imag\Sigma(\omega=0)$, [(e),(f)] the coherence
  scale $\Gamma^*/T$, and [(g),(h)] the resistivity $\rho$, all plotted
  as functions of temperature on  a linear (left) and 
a logarithmic (right) scale for \Mone\ (black), \Hone\ (yellow),
  \Itwo\ (red), and \Wzero\ (blue). Symbols are defined as in
  \Fig{fig:Tdep_1}. [(d),(h)] The dashed grey guide-to-the-eye lines
  indicate FL behavior. [(e),(f)] The horizontal dashed grey lines mark
  $\Gamma^*/T^*=1$. [(g),(h)] The horizontal solid 
grey line marks the MRI 
limit defined via $k_{\textrm{F}}l_{\textrm{min}}\approx2\pi$.}
\label{fig:Tdep_2}
\end{figure*}

\subsection{Scattering rate   at the Fermi level}
 
Figure \ref{fig:Tdep_2} shows the temperature dependence of the quasiparticle weight, the scattering rate,  the coherence scale,
 and the resistivity. We now discuss them in turn.

The scattering rate $-\imag\Sigma(\omega=0)$ is plotted as a function of temperature in  \Figs{fig:Tdep_2}(c) and \ref{fig:Tdep_2}(d). 
For $T<T_\fl$, $-\imag\Sigma(\omega=0)$ follows FL behavior [cf.\ dashed grey guide-to-the-eye line in  \Fig{fig:Tdep_2}(d)].
In \Hone{}, for $T>T_\fl$, the scattering rate is small and shows a crossover to a rather flat behavior in the HQP regime. By contrast, in \Mone{},
the scattering rate  increases strongly 
 [cf.\ \Fig{fig:Tdep_2}(e)], saturating at  high temperatures due to the presence of a pseudogap.
\Itwo{} shows a mixture of both the Hund and the Mott behavior. $-\imag\Sigma(\omega=0)$ first flattens somewhat for $T_\fl<T<\TspinO$, but then increases strongly for $T>\TspinO$, saturating as well at very high temperatures. Notably, $-\imag\Sigma(\omega=0)$ is larger for \Itwo{} than for \Mone{} for $T<0.1$; this is  caused by the larger $J=2$ in \Itwo{}.
The scattering rate in \Wzero{} is small and FL-like. It keeps growing slowly with increasing temperature.

\subsection{Coherence scale}
\label{subsec:coherencescale}

In \Figs{fig:Tdep_2}(e) and \ref{fig:Tdep_2}(f) we plot
 $\Gamma^*/T$, with the inverse QP lifetime, defined as 
\begin{align}
\Gamma^*(T)=-Z(T)\imag\Sigma(\omega=0,T) \, .
\end{align} 
In a FL, i.e., for $T\lesssim T_{\fl}$, one expects $\Gamma^*(T)\propto T^2$.
The coherence scale $T^*$ is defined as $\Gamma^*/T^*\equiv1$
(cf.\ intercepts with horizontal dashed grey line).  Above $T^*$
coherent Landau QPs become short-lived and the FL picture breaks down.

\Hone{} is characterized by a very broad maximum of $\Gamma^*/T$ in
the NFL regime around $\TspinO$. This behavior is reminiscent of DFT+DMFT
results for \sroone, where $\Gamma^*/T$ keeps increasing in a
FL-to-NFL crossover regime above $T^*\approx100$\,K and finally reaches a plateau above $350$\,K~\cite{Mravlje2011}. 
By contrast, \Mone{} shows only a narrow
plateau in $\Gamma^*/T$ around $\TspinO$ before it diverges [due to
  the divergence of $Z(T)$]. Again, \Itwo{} features a mixture of both
the Hund and the Mott behavior. $\Gamma^*/T$ first exhibits a maximum
at $\TspinO$, but then diverges
above $\TorbO$. In \Wzero, $\Gamma^*/T$
is very small and grows linearly
with increasing temperature, implying $\Gamma^\ast \propto T^2$.

\subsection{Resistivity}

The resistivity $\rho(T)$ is shown in \Figs{fig:Tdep_2}(g) and \ref{fig:Tdep_2}(h). In the
FL regime, we find $T^2$ behavior (though this is hard to resolve very
accurately).  Equivalently to the findings for 
a hole-doped Mott insulator~\cite{Deng2013}, we observe 
for \Hone\ and \Itwo\ that in the 
regime $T_\fl<T<\TspinO$, $\rho(T)$  first
  increases approximately linearly with a negative intercept, then it
  shows 
a slope-decreasing knee-like feature, above which
  a linear increase with positive intercept sets in.
  The inset of  \Fig{fig:Tdep_2}(g) highlights this for I2  using grey dashed lines, which approximate the behavior of the red curve.
   For \Hone\ (yellow
    curve), $\rho(T)$ keeps increasing linearly up to the highest
  temperature plotted, and thus behaves qualitatively in the same way
  as the hole-doped Mott insulator of \oRef{Deng2013}. This is an
  intriguing similarity, considering that both systems are assumed to
  be governed by resilient QPs in their NFL regime. Moreover, our findings for
  \Hone{} are reminiscent of the DFT+DMFT simulations~\cite{Haule2009}
  and measurements~\cite{Hardy2013} of the resistivity in iron
  pnictides.  In contrast to \Hone, for \Itwo\ a second
  (slope-increasing) knee occurs at $\TspinO$, beyond which
  $\rho(T)$ grows rapidly with increasing temperature until it
  saturates above $\TorbO$ in the presence of a stable pseudogap. For
  \Mone\ (black curve), we do not observe a
slope-decreasing knee, but  instead a slope-increasing knee
at $T\approx0.08$, above which
  $\rho(T)$ increases rapidly with growing temperature
[cf.\ \Fig{fig:Tdep_2}(h)]. \Wzero{} is
  again characterized by a large FL  regime, reaching up to
  very high temperatures. For all but the
  largest temperatures, 
  $\rho(T)$ is much smaller for the system 
  with $J=0$  than for those with finite $J$. (At very high $T$,
  the resistivity $\rho(T)$  of \Wzero\ increases past that of 
  \Hone; the reason is that the scattering
  rate $-\mathrm{Im}\Sigma(\omega=0,T)$ of \Wzero\ likewise
  increases past that of \Hone\ [cf.~Figs.~\ref{fig:Tdep_2}(c) and \ref{fig:Tdep_2}(d)], reflecting the   fact that the former
  has a larger bare gap,  
  $\Delta_b^{\Wzero} = 3.5$ vs
  $\Delta_b^{\Hone} = 1$.)   We remark that for all systems $\rho(T)$
  crosses the Mott-Ioffe-Regel (MIR) limit, $\rho_\MIR$  
[cf.\ horizontal solid
 grey line in \Figs{fig:Tdep_2}(g) and \ref{fig:Tdep_2}(h) and \App{sec:appendix_res} for a definition of $\rho_\MIR$] 
and   continues to grow above this limit. 
 As expected, \Mone{} crosses the MIR limit at a
  smaller temperature scale than \Hone{}. Notably, \Itwo{} crosses the
  MIR limit at an even lower scale although Coulomb interactions are
  larger in \Mone{} than in \Itwo{}. This strong correlation effect is
  due to Hundness, i.e., large $J$. 

To conclude this subsection, we remark that an analysis of the temperature dependence of the optical conductivity $\sigma(\omega)$ for 
\Itwo{} is presented in \App{sec:TdepQP}.

\begin{figure*}
\centering
\includegraphics[width=0.8\linewidth, trim=14mm 90mm 14mm 41mm, clip=true]{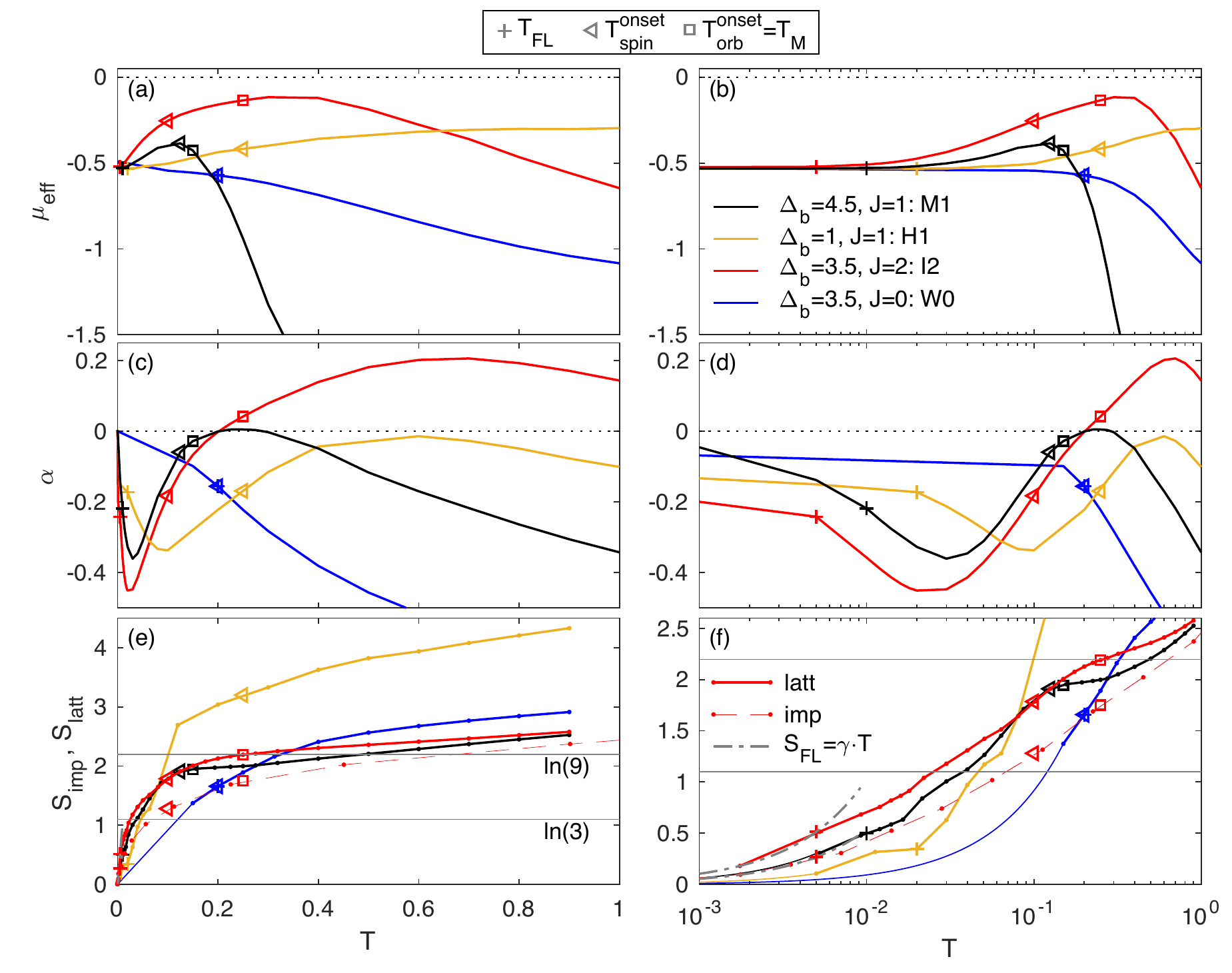}
\caption{[(a),(b)] The effective chemical potential $\mu_\eff$, [(c),(d)] the
  thermopower $\alpha$, and [(e),(f)] the lattice
  entropy $S_\latt$ (solid)  and the impurity
  contribution to the entropy $S_\imp$ (dashed), all plotted as functions of temperature on a
  linear (left) and  logarithmic (right) scale for \Mone{}
  (black), \Hone{} (yellow), \Itwo{} (red), and \Wzero{}
  (blue). Symbols are defined as in \Fig{fig:Tdep_1}. In (f), the grey
  dash-dotted curves indicate FL behavior for $S_\imp$ and $S_\latt$,
  respectively. We remark that wiggles in $S_\latt$ are an artefact due to few data points used in its computation.}
\label{fig:Tdep_3}
\end{figure*}

\subsection{Effective chemical potential of quasiparticles}

We now turn to \Fig{fig:Tdep_3}. We first study the evolution of the
effective chemical potential for  QPs,
$\mu_\eff=\mu-\real\Sigma(\omega=0)$, in
\Figs{fig:Tdep_3}(a) and \ref{fig:Tdep_3}(b). For $T<T_\fl$, $\mu_\eff$ is constant, i.e.,
Luttinger pinning holds (cf.\ Sec.~3.10.2 of \oRef{Stadler2019} for
details). Interestingly, for the finite-$J$ systems $\mu_\eff$
increases towards $0$ with increasing temperature, $T_\fl<T<\TspinO$,
i.e., towards an effective half-filling of the system. In \Hone{}, this
trend is retained above $\TspinO$ until $\mu_\eff$ approaches a
plateau in the mixed-valence regime. This behavior fits to the SOS
screening picture (cf.\ \Figs{fig:scrsketch} and \ref{fig:scrsketchT})
where, above $T_\fl$, spins are gradually unscreened to form an
effective ${3}/{2}$ spin (which implies effective half filling), while the orbitals are still in an orbital
singlet for $T<\TspinO$. For $T>\TspinO$, the orbitals start to get
unscreened while large quasi-free spins persist. In \Mone{},
$\mu_\eff$ drastically reduces for $T>T_{\textrm{M}}$,
reflecting the formation of a pseudogap. In \Itwo{},
$\mu_\eff$ first increases markedly almost up to $0$ and then
decreases for $T>\TorbO$, similarly to \Mone{}. By contrast, for $J=0$,
\Wzero{} directly decreases above $T_{\fl}$. The
substantial continuous increase of $\mu_\eff(T)$ with increasing
temperature towards half-filling, i.e., an inflating Fermi volume, is
clearly connected to the existence of a finite $J$ in the \HHM, while the
decrease of $\mu_\eff(T)$ with increasing temperature is a Mott
feature.

\subsection{Thermopower}

In \Figs{fig:Tdep_3}(c) and \ref{fig:Tdep_3}(d) we show the thermopower (Seebeck
coefficient) $\alpha(T)$ [as defined in \Eq{eq:alpha}] and compare
the \HHM results to the thermopower of {\sroone} reported in
\oRef{Mravlje2016}. In the FL regime, the thermopower of the \HHM at
$n_d=2$ shows an electron-like decrease, i.e., $\alpha(T)<0$. 
This is qualitatively consistent (modulo a particle-hole
transformation) with the hole-like increase, $\alpha(T)>0$, 
observed for {\sroone}, which in a \HHM-type description
would correspond to $n_d=4$. However, our
data is not accurate and dense enough to unveil FL behavior,
$\alpha(T)\propto T$. Similar to the (broad) maximum in $\alpha(T)$ of
{\sroone} around 300\,--\,500\,K, we observe a minimum
in the crossover regime $T_\fl<T<\TspinO$. In \Hone{}, we further find
a saturation (broad maximum) well above $\TspinO$. In \Itwo{} and
\Mone{}, a maximum occurs above $\TorbO$, as well.  Overall, the
behavior of $\alpha(T)$ is similar for all systems with finite
$J$. However, the minimum is much more extended and lies at higher
energies in \Hone{} compared to \Mone{} [cf.\ \Fig{fig:Tdep_3}(c)]. In
contrast, \Wzero{} with $J=0$ does not exhibit any minimum (or
maximum) in $\alpha(T)$. Here, the thermopower decreases in a FL-like
fashion in an extended temperature range.

In sum, we conclude that \Hone{} reflects the findings of
\oRef{Mravlje2016}. Using $t\approx5000$ K (a value which is estimated
from a comparison of the model bandwidth with the realistic bandwidth
of {\sroone}~\cite{Deng2019}), the minimum of $\alpha(T)$ of \Hone{}
is indeed in the same temperature range
(300\,--\,500\,K) as the maximum observed for
{\sroone}. Our results support the suggestion made in  \oRef{Mravlje2016} that this
unusual feature in $\alpha(T)$ can be associated with quenched
orbitals and fluctuating spins as present in the two-stage SOS
screening process. To be more precise, the minimum of $\alpha(T)$ in
the \HHM corresponds to the crossover regime, where the spins get
gradually screened to form coherent Landau QPs. Thus, this minimum in
$\alpha(T)$ is observed together with the formation of the step-like
\ARPESt{} feature [cf.\ \Fig{fig:AkwDeltab1J1}(d)]. 

\subsection{Entropy}
\label{sec:entropy}
We conclude our study of Hund and Mott features in the \HHM\ by
calculating the lattice entropy for  \Hone{}, \Mone{},  \Itwo{}, and \Wzero{}. For  \Itwo{}, 
we additionally calculate  the impurity contribution to the
entropy  [cf.\ \Fig{fig:Tdep_3}(e) and \ref{fig:Tdep_3}(f)]. We start our discussion with \Itwo{}.
For the computation of the lattice entropy $S_{\latt}(T)$, we use
\Eq{eq:Slatt}.  The impurity contribution to the entropy
$S_{\imp}(T)$ is obtained with \Eq{eq:Simp}.  
Remarkably, we find that
$S_{\latt}(T)$ is larger than $S_{\imp}(T)$ in the whole temperature
range $0<T<1$, while both entropies behave qualitatively in the same
way. The difference between $S_{\latt}(T)$ and $S_{\imp}(T)$ already
arises in the FL regime, where the entropy is given as 
\begin{subequations}
\label{eq:entropy}
\begin{eqnarray}
S(T)&=&\gamma T \, \\
\gamma&=& \frac{2N_c \pi^2}{3\mathcal Z}.
\end{eqnarray}
\end{subequations}
When computing the lattice or impurity entropies, $S_\latt$ or $S_\imp$, the parameter $\mathcal{Z}$ should be equated to the mass renormalizations,
$Z_\latt$ or $Z_\imp$, derived from the lattice or impurity Green's functions, respectively. The former is given by  $Z_\latt=
[1-\left. \partial_{\omega} \real \Sigma(\omega)\right|_{\omega=0}]^{-1}$.
The latter, found by a first-order expansion of $G_{\imp}(\omega)=
[\omega-\varepsilon_d-\Delta(\omega)-\Sigma(\omega)]^{-1}$,  
where $\Delta(\omega)$ is the self-consistent hybridization function,
is given by $Z_\imp =
[Z_{\latt}^{-1}-\left. \partial_{\omega} \real\Delta(\omega)\right|_{\omega=0}]^{-1}$  
(cf.\ Sec.~3.9 in \oRef{Stadler2019} for details). 
Obviously,  DMFT generically yields $Z_\latt < Z_\imp$ in the FL regime (when using a Bethe lattice). This implies that $S_\latt > S_\imp$, as found numerically
above. Although
this insight can be simply derived, we are not aware of any previous
results that explicitly demonstrated this quantitative difference of
the impurity and the lattice entropy. Its implication is that $S_\imp$ can not be regarded as a quantitatively reliable proxy for $S_\latt$.

Nevertheless both entropies for \Itwo{}   reveal the two-stage SOS screening process. For $T>\TorbO$, \Itwo{} is characterized by a pseudogap and both the spin and orbital degrees of freedom are unscreened,  resulting in $S_\latt>\ln(9)$. [$S_\latt$ slightly exceeds $\ln(9)$ because of remaining active charge fluctuations in the pseudogap regime.]
 $S_{\imp}$ crosses $\ln(9)$ at slightly higher temperatures.  For $T<\TorbO$, $S_\latt(T)$ and $S_\imp(T)$ decrease continuously with decreasing temperature, reflecting the screening of orbital degrees of freedom, while spin degrees of freedom are still quasi-free. We observe that $S_\latt(T)$ crosses $\ln(3)$ below $\TspinO$, while $S_\imp(T)$ crosses $\ln(3)$ at about $\TspinO$. The value $\ln(3)$ is associated with a spin triplet and an orbital singlet.  For $T<T_\fl$ we find  FL behavior for both  $S_\latt$ and  $S_\imp$, indicated by the dash-dotted grey fits, respectively [cf.\ \Fig{fig:AkwDeltab1J1}(f)]. 

Overall, we clearly observe that the two-stage SOS screening process is a continuous process: the entropy continuously decreases with decreasing temperature, i.e., no stable NFL fixed point is reached in the system (this was already pointed out in the Supplemental Material of \oRef{Stadler2015}). Instead, we are faced with an intriguingly complex crossover behavior.

The two-stage SOS  screening process is also  manifest in $S_\latt$ for \Hone{} and \Mone{}. While the qualitative behavior is similar, quantitative details differ.  In the FL regime,  $S_\latt$ is smaller for \Hone{} than for \Mone{}, since $S_\latt\propto T/Z_\latt$ (and \Hone{}, having smaller $U$, has less mass enhancement, i.e., larger $Z_\latt$). Above $T_\fl$,   $S_\latt$ increases strongly for \Hone{}, leading to a very large entropy  ($>\ln(9)$) above $\TspinO$. We interprete this as a consequence of large charge fluctuations due to small Coulomb interactions. By contrast, $S_\latt$ for \Mone{} approaches $\ln(9)$ above $\TM$ and only slightly exceeds $\ln(9)$ for very high temperatures. 

Very recently, a detailed study of the temperature dependence of the entropy and specific heat of a three-band Hubbard model has been performed \cite{Yue2020}. This study is much more comprehensive than ours. Their results are not directly comparable to ours, although, since their interaction term contained only density-density terms but no spin-flip terms.

\section{Conclusion} 
\label{sec:Concl}

\begin{figure*}
\centering
\includegraphics[width=0.9\linewidth, trim=0mm 0mm 0mm 0mm, clip=true]{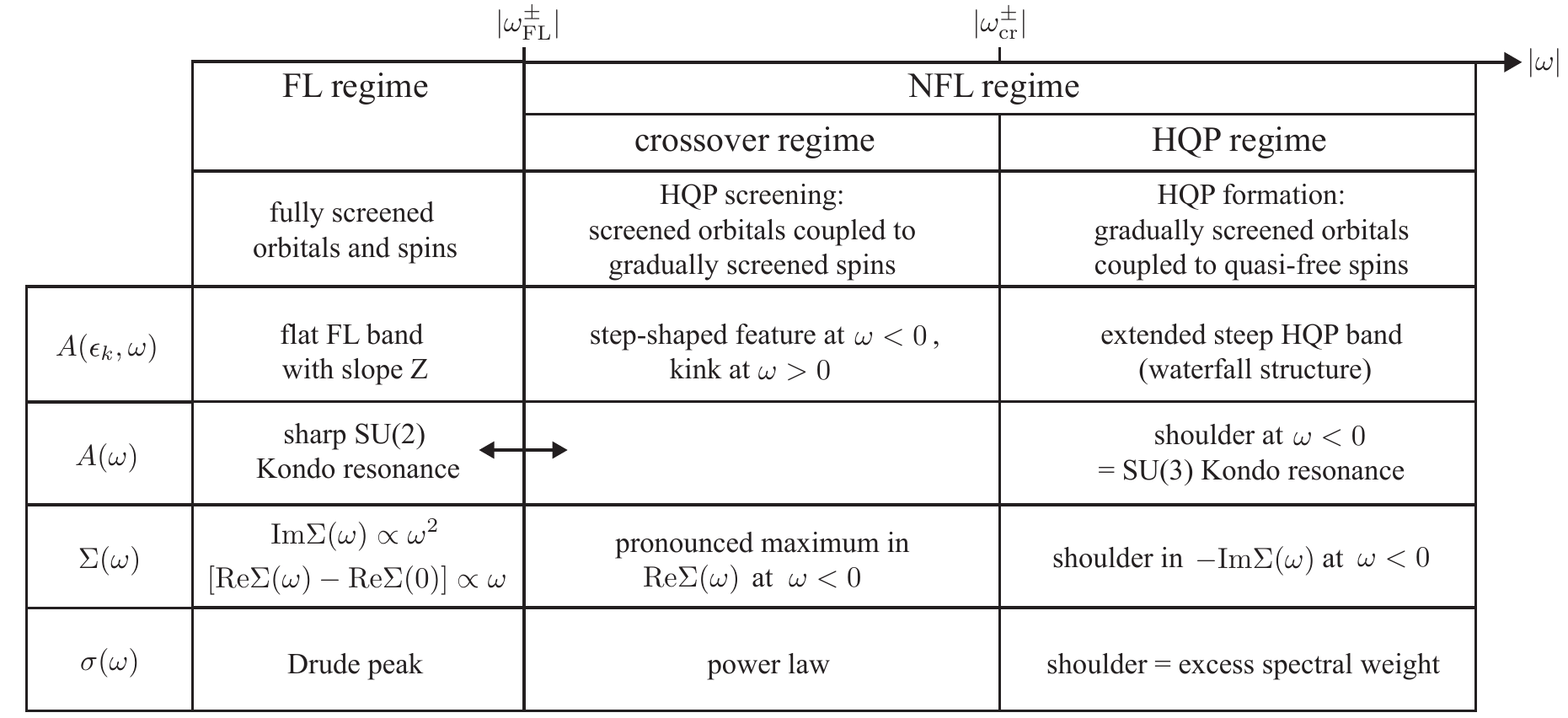}
\caption{Overview of important SOS features in the \HHM for $n_d=2$ at $T=0$. Features are described as functions of decreasing frequency.
}
\label{fig:sum1}
\end{figure*}

\subsection{Fingerprints of Hund versus Mott physics}

In this paper we have used DMFT+NRG to investigate the normal state properties of the  degenerate three-band Hubbard-Hund model (\HHM) 
 with focus on $1/3$ filling, a minimal model with relevance for Hund metals. Our paper has been based on the following key question: 
What are the decisive fingerprints of a Hund metal as opposed to a Mott-correlated metal? We conclude by giving a summary-style overview of the fingerprints found in the present paper.

At $T=0$, finite $J$ induces an intertwined two-stage SOS Kondo-type screening
process in the \HHM at $n_d=2$, in which orbital and spin
   degrees of freedom are explicitly coupled:  below 
 $\Torb$, the orbital degrees of freedom form an orbital singlet through
 the formation of a large effective Hund's-coupling-induced
impurity spin of ${3}/{2}$---\textit{including} a bath spin degree of freedom; and 
below  $\Tspin$, the spin-$3/2$ is fully screened by the three bath channels of the \HHM. In the frequency domain this screening process
 results in three characteristic
regimes: a FL regime, a NFL crossover regime, and a NFL HQP
regime. 
At zero temperature, clear signatures of SOS include: (i) a low-frequency  FL regime with a narrow ``needle''-formed SU(2) Kondo peak in the local density of states, a low-frequency Landau QP band with a small  slope given by $Z$ in \ARPESt{} spectra, FL scaling of the self-energy, a Drude peak in the optical conductivity (cf.\ \App{sec:TdepQP}); (ii) a NFL crossover regime signaling the deviation from FL behavior characterized by a step-like feature  in  the dispersion at $\omega<0$ and a kink at $\omega>0$ [accordingly, $\real\Sigma(\omega<0)$ exhibits a pronounced maximum]; and (iii) an intermediate-frequency NFL ``Hund quasiparticle'' (HQP) regime with a SU(3) Kondo resonance in the local density of states, also identifiable as excess spectral weight in the optical conductivity (cf.\ 
\App{sec:TdepQP}) and as  a resilient slightly particle-hole asymmetric steep ``HQP band'' in \ARPESt{} spectra (waterfall structure), which is extended over a large frequency range, where the scattering rate is   only weakly energy dependent  [e.g., there is a shoulder in $\imag\Sigma(\omega<0)$]. We remark that the particle-hole asymmetry of the \HHM leads to two distinct FL scales in the frequency domain and to very different features in the SOS window at negative and positive frequencies (e.g., in \ARPESt{}
spectra).  
These SOS features (cf.\ \Fig{fig:sum1} for an overview) 
are generic and are found for both the
metallic \Hone\ and the metallic \Mone, since SOS physics is
essentially impurity physics~\cite{Stadler2015}. However, there is an
important difference. 

A Hund metal, such as \Hone, lies far from any MIT phase
boundary. Strong correlations are primarily induced by the two stage SOS Kondo-type
screening, which leads to the localization of spins rather than
charges.  The incoherent SOS window is extended over a broad range of
energies, reaching up to bare excitation scales. In the \HHM{}, at high
frequencies, the SU(3) Kondo resonance (shoulder) merges with the
Hubbard bands. At very low temperatures, the local density of states
exhibits a two-tier quasiparticle peak on top of a broad incoherent
background.

By contrast, Mott-correlated metals with $\sim1/3$ filling such as {\vo} \cite{Deng2019}, represented in our study by \Mone, 
are close to the MIT phase boundary. Thus, at zero temperature, both $\Torb$ and $\Tspin$ are strongly reduced compared to bare excitation scales and the SOS window is very small, i.e., a narrow QP peak exists between well-separated pronounced Hubbard bands.  

\begin{figure*}
\centering
\includegraphics[width=0.9\linewidth, trim=0mm 0mm 0mm 0mm, clip=true]{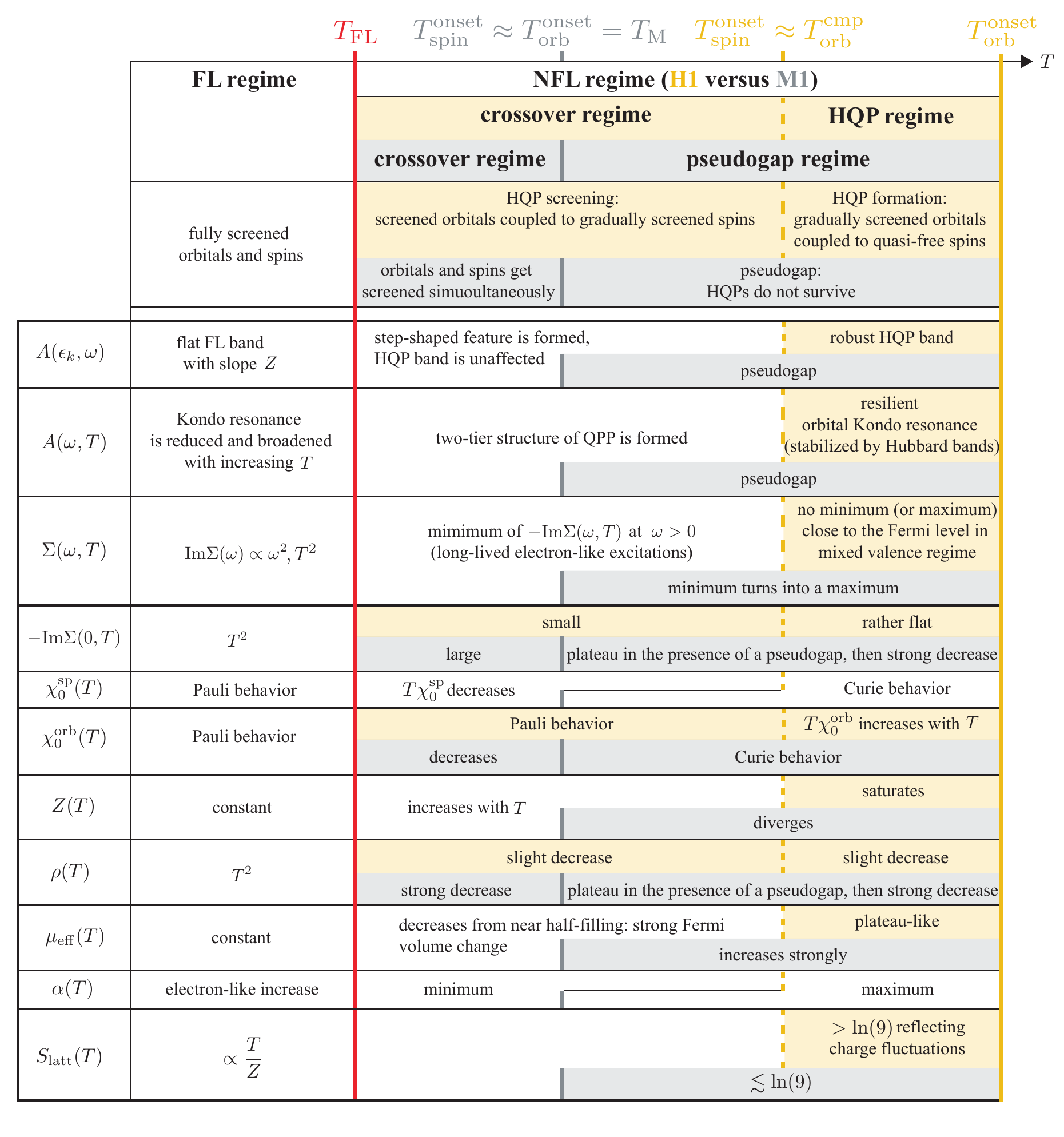}
\caption{Overview of important Hund and Mott features in the temperature dependence of various physical quantities. Hund-related features are marked yellow, Mott-related features are marked grey. Common features are on white background. Note that the temperature scale is only schematic.
Features are described as functions of decreasing temperature.}
\label{fig:sum2}
\end{figure*}

In Hund metals, the SOS screening process also governs the temperature
dependence of Hund metals, up to highest temperatures. Most
importantly, we argue that the nature of the incoherent transport
regime is governed by resilient HQPs, while the FL regime is described
in terms of Landau QPs. 
In \oRef{Deng2019}, we have identified two
different temperature scales for the onset of orbital and spin
screening in Hund metals, $\TorbO$ and $\TspinO$, respectively.  For
$\TspinO <T<\TorbO$, HQPs dominate the high-temperature physics and
lead to a Curie-like static spin susceptibility (while the static
orbital susceptibility is a decreasing function of temperature) and a
resilient QP peak (without substructure) in the local density of
states. In the \HHM, we find a robust HQP band in \ARPESt{} spectra, an
additional HQP plateau in $Z(T)$, a rather flat (electron-like)
scattering rate, a linear  resistivity exceeding the MIR limit, and an
inflated Fermi volume ($\mu_{\rm eff}$ increases with increasing temperature).
 At very high temperatures, mixed-valence
physics additionally comes into play.  Due to the special SOS
screening process, the spins can only get screened as soon as the
orbitals are fully screened at $T^{\cmp}_{\orb}$, thus $\TspinO\approx
T^{\cmp}_{\orb}$. For $T_{\fl} < T < \TspinO$ also the spins are  gradually
screened, eventually resulting in the full screening of both degrees
of freedom and thus in a FL below $T_{\fl}=\TspinC$. The spin screening is
signalled by the formation of a step-like feature in ARPES spectra,
while the completion of orbital screening is characterized by a
Pauli-like orbital susceptibility. In this regime, the thermopower has
a minimum. A corresponding feature in the thermopower is observed in experiments for ruthenates.
\cite{Mravlje2016}.

By contrast, in Mott-correlated metals, with increasing temperature, SOS features (and HQPs) only survive at very low temperatures, whereas the behavior at higher temperatures is fully governed by classical Mott physics (as known from the one-band Hubbard model): the DMFT self-consistency condition opens up  a pseudogap in the local spectrum  by localizing the charges. Conversely, with decreasing temperature,  spin, and orbital degrees of freedom  get screened simultaneously at the temperature scale, $T_{\textrm{M}}=\TorbO\approx \TspinO$, 
 with the onset of a Kondo resonance, driven by DMFT. Only below $T_\fl=\TspinC\approx \TorbC$ both the spin and the orbital degrees of freedom get fully screened.

All important temperature-dependent signatures for \Hone{} and \Mone{} are summarized in \Fig{fig:sum2}. 

In sum, we shed light on two qualitatively different screening routes from the atomic degrees of freedom to the emerging heavy QPs in strongly correlated systems, driven by Hundness or Mottness, and corroborated that Hundness, i.e., SOS Kondo-type screening, dominates the anomalous physics of Hund metals in terms of resilient HQPs.

\subsection{Physics beyond the minimal three-band Hund-Hubbard model}

In the present study we purposefully focused on the 
\HHM, the simplest possible Hamiltonian capturing the essence of Hund and Mott physics. We thereby neglected several complications occurring in real materials.
Let us now briefly comment on these.
First, to fully exploit the power of the NRG,  we used a Coulomb interaction matrix with U(1)${}_\charge\times$SU(2)${}_\spin\times$SU(3)${}_\orb$ symmetry,  avoiding more realistic parametrizations of the Coulomb interaction such as the Kanamori parametrization.
Second, we neglected the spin-orbit coupling, which reduces the symmetry to U(1)${}_\charge\times$SU(2)${}_\textrm{tot}$ or even weaker symmetries, where ``tot'' stands for total angular momentum. 
The spin-orbit coupling terms have been shown to be irrelevant in the renormalization group sense~\cite{Horvat2017}, i.e., they do not affect the system's low-energy behavior unless the coupling strength is larger than $\Torb$.
Third, we neglected crystal field splittings.
Fourth, we took a very simple bipartite Bethe lattice, thereby ignoring effects arising from realistic electronic dispersions and Fermi surfaces. Spin-orbit coupling, crystal fields, and realistic band structures all bring about important physical effects not present in our model.
These include orbital  differentiation and even orbital-selective Mott transitions (see, for example, Refs.~\cite{deMedici2009,Vojta2010,deMedici2005,Ferrero2005,Yi2013}),
where one orbital becomes  much more correlated than others or even completely localized. Incorporating such realistic aspects 
is the focus of intensive current investigations in multiple materials and models (see, for example, Refs.~\cite{Bramberger2021,Kim2017,Springer2020}). Such studies 
will benefit from the deeper understanding, achieved
 in our paper, of the finite-temperature Hund metal state and how it is modified as the Mott transition is approached. In this sense, our
paper sets the stage for future studies incorporating additional material-specific physical effects.

Finally, an important aspect that was not studied in our paper
is the appearance of symmetry-broken  phases
in Hund metals at low temperatures, e.g., magnetic~\cite{Hoshino2016,Alloul2016},
insulating~\cite{Isidori2019}, and superconducting~\cite{THLee2018,Werner2016,Hoshino2015,Fanfarillo2020} phases. 
Generalizations and extensions of the
DMFT+NRG approach used here could be developed to achieve a deeper understanding of these phases, and how they emerge from the Hund metal state.

In the long run, such studies would also have to include the effects of nonlocal correlations and nonlocal interactions, neglected here, e.g., by using nonlocal extensions of DMFT \cite{Maier2005,Rohringer2018,Ayral2015,Ayral2016,Toschi2007,Held2008,Rubtsov2008,Brener2008,Taranto2014,Vilardi2018,Vilardi2019}.
Nonlocal correlations are generally expected to be weaker in Hund metals than Mott systems~\cite{Semon2017}.
We also expect nonlocal interactions to be less important as the screening of the nonlocal interactions is more efficient in metallic systems. 
Nevertheless, clarifying how nonlocal correlations and nonlocal interactions affect the physics of Hund metals is a very interesting question which is only beginning to be studied~\cite{Ryee2020}.

\subsection{Experimental signatures of two-stage screening}

Although our minimal \HHM\ neglects 
numerous effects relevant for realistic materials, as discussed above, the physics, which 
it does capture, in particular two-stage screening and SOS, is expected to be robust. Indeed, indications
of two-stage
screening of electrons have been found in several
experimental studies.
For example, they were
identified in various members of the iron pnictides and chalcogendies by means of infrared spectroscopy 
\cite{Schafgans2012,Yang2017}; resistivity, heat-capacity, thermal-expansion,  susceptibility measurements \cite{Hardy2013,Hardy2016};
quasiparticle scattering interference~\cite{Kostin2018}; proximity effect~\cite{Song2020};
and ARPES \cite{Miao2014,Miao2016}.  
A second prototypical system of a Hund metal  is {\sroone},  
where optical conductivity \cite{Stricker2014}, thermopower
\cite{Mravlje2016}, and ARPES \cite{Tamai2019} provide multiple signatures of Hund metal behavior. We hope that the present paper of a minimal 
three-band model, containing the minimal ingredients to yield Hund and/or Mott physics, will assist future experimental studies in attributing observed features to either Hund rule effects (Hundness) or charge-blocking effects (Mottness).

\acknowledgments
We thank F.~B.~Kugler for helpful discussions.
K.M.S., S.-S.B.L., and J.v.D.~were supported by the Deutsche Forschungsgemeinschaft (DFG, German Research Foundation) under Germany's Excellence Strategy EXC-4 (Project No.~24040814) and EXC-2111 (Project No.~390814868) and through Project No.~409562408. S.-S.B.L.~acknowledges the DFG grant LE3883/2-1 (Project No.~403832751).
A.W. was supported
by the U.S. Department of Energy, Office of Science, Basic Energy Sciences, Materials Sciences and Engineering Division
under Contract No. DE-SC0012704.
G.K. was supported by NSF grant DMR-1733071.

\appendix
\section{Asymmetry of frequency-dependent quantities at zero temperature}
\label{sec:asymfdq}

\begin{figure*}
\centering
\includegraphics[width=0.9\linewidth, trim=14mm 50mm 14mm 0mm, clip=true]{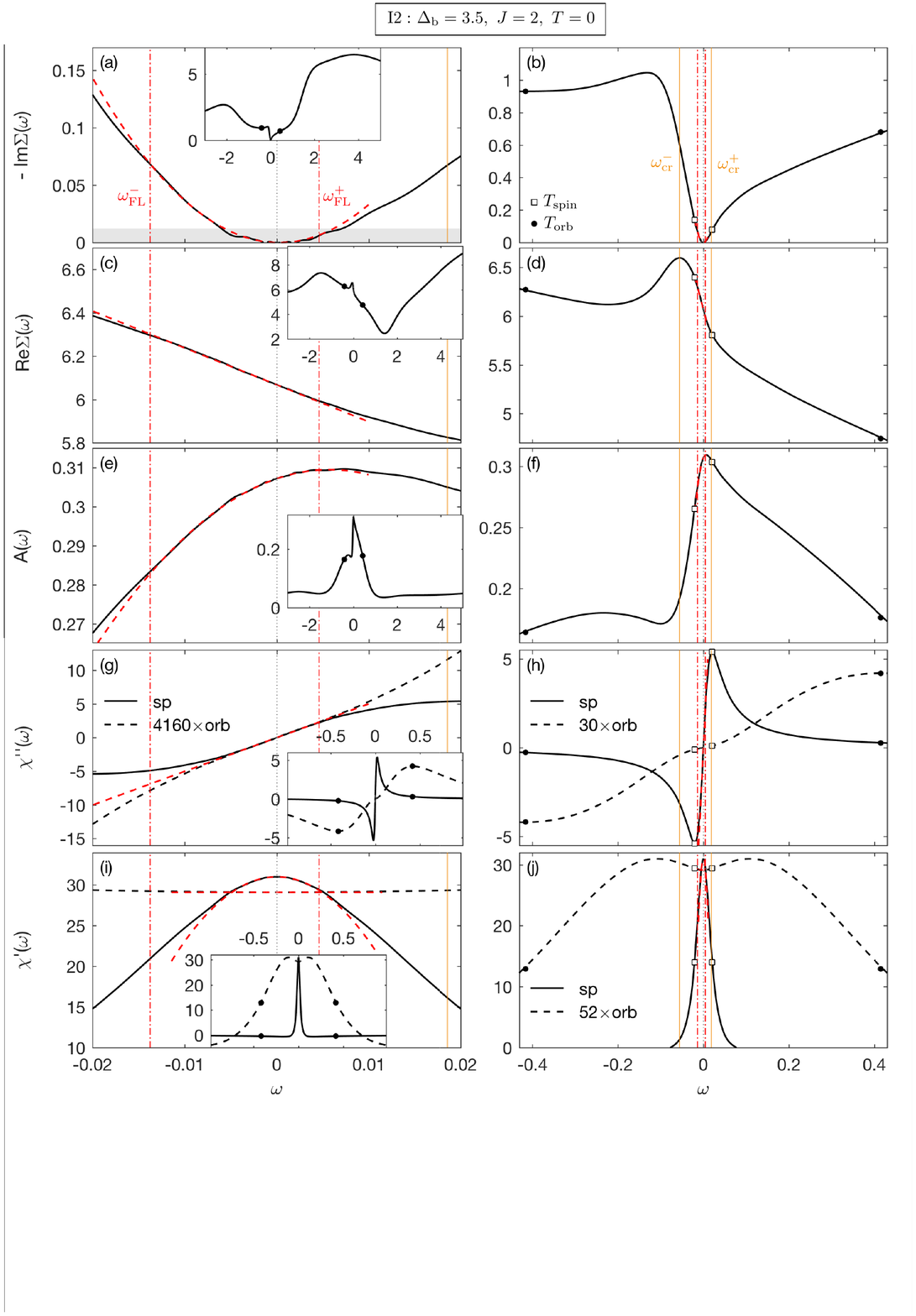}
\caption{
 [(a),(b)] The imaginary part $\imag\Sigma(\omega)$ and [(c),(d)] the real part $\real\Sigma(\omega)$ of the self-energy; 
 [(e),(f)] the local spectral function $A(\omega)$; [(g),(h)]
the imaginary part $\chi''(\omega)$ and  [(i),(j)]  the real part $\chi'(\omega)$
of the spin (solid) and orbital (dashed) susceptibilities are plotted versus frequency 
for \Itwo{} ($\Delta_b=3.5$, $J=2$) at $T=0$. Left panels are zooms into the FL regime, 
whereas their insets show the quantities on a large frequency range.
The SOS window is presented in the right panels.
Dashed red fits reveal FL behavior for $\imag\Sigma(\omega)$,
 $\real\Sigma(\omega)$, and $A(\omega)$
in the asymmetric range, $\omega^{-}_{\fl}<\omega<\omega^{+}_{\fl}$, with $\omega^{+}_{\fl}\approx\tfrac{1}{3}\omega^{-}_{\fl}$
(indicated by vertical dash-dotted red lines) and for the orbital and spin susceptibilities in the symmetric range, $|\omega|<\omega^{+}_{\fl}$.
The vertical solid yellow line at $\omega<0$ denotes the energy scale $\omega^{-}_\cross$ of the maximum in $\real\Sigma(\omega)$ at $\omega<0$. In (b), $\omega^{+}_\cross=-\tfrac{1}{3}\omega^{-}_\cross$ marks the kink in $\real\Sigma(\omega)$ at $\omega>0$. Filled dots and open squares mark the orbital and spin Kondo scales, respectively. The grey area in (a) indicates a systematic error in $\imag\Sigma(\omega)$ (cf.\ Sec.~3.2 of \oRef{Stadler2019} for details).}
\label{fig:asym_1} 
\end{figure*}

In this Appendix,  we investigate in more detail the particle-hole asymmetry of the \HHM{} 
at zero temperature discussed in the main text. In particular, we look at the frequency-dependence 
of the self-energy, the local spectral function, the dynamical spin and orbital susceptibilities, the optical conductivity, and the kinetic energy.

A first detailed temperature-dependent study of the implications of
particle-hole asymmetry in Hubbard-type models was given in
\oRef{Deng2013} for a one-band hole-doped Mott insulator, i.e., for a
model with only one type of degrees of freedom (spins). It was shown
that a well-defined QP peak of ``resilient'' QP excitations exists above the FL scale $T_{\fl}$ and that it dominates an
intermediate incoherent transport regime up to
$T_\MIR$. Above this temperature the resistivity exceeds the
MIR limit (cf.\ \App{sec:calc_optcond} for a definition) and the resilient QPs
eventually disappear, or more specifically, the QP peak merges with
the lower Hubbard band. Interestingly, the resilient QPs are longer-lived for
electron-like than for hole-like excitations, due to the particle-hole
asymmetry in the model. This asymmetry further leads to different
scales, $\omega^{-}_{\fl}$ and $\omega^{+}_{\fl}$, below which FL
behavior is found at negative and positive frequencies at $T=0$.

In \Fig{fig:asym_1} we revisit the self-energy $\Sigma(\omega)$, the
spectral function $A(\omega)$, and the orbital and spin
susceptibilities, $\chi_\orb(\omega)$ 
and $\chi_\spin(\omega)$ [\Eqs{subeq:susceptibility}], at
$T=0$. We consider system \Itwo\ ($\Delta_b=U-2J=3.5$, $J=2$)
which features a broad SOS window, well separated from the Hubbard side bands. 
We start with a detailed investigation of the FL
regime (cf.\ left panels of \Fig{fig:asym_1}) and then concentrate on
the SOS window (cf.\ right panels of \Fig{fig:asym_1}). Due to the
universal behavior of the model with respect to $\Delta_b$
(respectively $U$) (cf.\ Fig.~10 of \oRef{Stadler2018}) the following
findings are generic in the metallic regime of the \HHM, but can occur
on very different energy scales (depending on the value of
$\Delta_b$).

\paragraph*{Asymmetry in the FL regime.}
\label{appendix:asym}
The left panels of \Fig{fig:asym_1} zoom into the frequency regime below $\Tspin$ (marked by open squares in the right panels). 
Similar to the results of \oRef{Deng2013} we observe  in \Figs{fig:asym_1}(a), \ref{fig:asym_1}(c), and \ref{fig:asym_1}(e) that FL behavior holds up to different frequency scales, $\omega^{-}_{\fl}$ and $\omega^{+}_{\fl}$, at $\omega<0$ and $\omega>0$ (cf.\ vertical red dash-dotted lines), respectively. These \fl{} scales have been identified in $A(\epsilon_k, \omega)$ in the main text. The FL behavior is indicated by the red dashed curves in \Figs{fig:asym_1}(a), \ref{fig:asym_1}(c), and \ref{fig:asym_1}(e): a parabola for $-\imag\Sigma(\omega)$ in panel (a), a linear fit for $\real\Sigma(\omega)$ in panel (c), and a  parabola for $A(\omega)$ in panel (e). Clearly, the black DMFT+NRG results  deviate earlier from the red FL curves  on the positive frequency side, i.e., at a lower scale $\omega^{+}_{\fl}\approx\tfrac{1}{3}\omega^{-}_{\fl}$. 
Furthermore, we find that the position of the maximum of $A(\omega)$ approximately coincides with $\omega^{+}_{\fl}$.  

In \Figs{fig:asym_1}(g) and \ref{fig:asym_1}(i) we show the imaginary and the real parts of the dynamical orbital and spin susceptibilities, $\chi_{\orb}(\omega)$ and $\chi_{\spin}(\omega)$ [cf.\ \Eq{subeq:susceptibility}], respectively. 
The imaginary part of the dynamical susceptibility is defined as $\chi''(\omega)\equiv-\tfrac{1}{\pi}\imag{\chi(\omega)}$, the real part as $\chi'(\omega)\equiv\real{\chi(\omega)}$.
In contrast to $\Sigma(\omega)$ and $A(\omega)$ these quantities are particle-hole symmetric. The imaginary parts of both the orbital and spin susceptibilities follow the red dashed linear  FL fit only for $|\omega|\lesssim\omega^{+}_{\fl}$. Accordingly, the real part of the spin susceptibility $\chi'_{\spin}(\omega)$  also exhibits parabolic FL scaling in this regime, while  the real part of the orbital susceptibility $\chi'_{\orb}(\omega)$ is essentially constant. 

In this paper we define the orbital and spin Kondo scales, $\Torb$ and $\Tspin$ (cf.\ open squares and filled circles in \Fig{fig:asym_1}), below
which Kondo screening of the local orbital or spin degrees of freedom sets in, as the peak positions of $\chi''_{\orb}(\omega)$ and $\chi''_{\spin}(\omega)$, respectively. As usual for crossover scales, other definitions
are possible, which would differ from ours by constant prefactors.

\section{On the definition of crossover scales}
\label{app:CrossoverScales}

Unlike  a phase transition occurring at a  well-defined critical  temperature, spin screening is  a crossover phenomenon, which  cannot be described in terms of  just a  single number.  This was understood very early in the classic work of K.~Wilson \cite{Wilson1975}.
To set the stage for the discussion of the Hund-Mott problem discussed in the main text, we here summarize some of Wilson's results for
the temperature dependence of the impurity contribution to the spin susceptibility $\chi_\mathrm{imp}(T)$.
(For a detailed discussion, see Section 
IX of Ref.~\cite{Wilson1975} or Section 4.6 in Hewson's book \cite{Hewson1993}.)

 Wilson studied the single-impurity Kondo model, 
 involving a single spin-$\frac{1}{2}$ impurity coupled to a 
 conduction band with a featureless (flat) density of state. 
 He considered the weak-coupling limit,  where the 
 impurity-bath exchange coupling $\Jk$
 is much smaller than the bandwidth $W$.
  He showed that in this limit the temperature dependence of physical quantities can be described  in terms of a crossover scale, the Kondo temperature $\Tk$, and  a universal scaling function,   $F(T/\Tk)$. For example $\chi_\mathrm{imp}(T)$ has the form \cite{Yuval1970,Anderson1970,Anderson1970b}    
 \begin{align}
 \label{eq:ScalingFunction}
\chi_\mathrm{imp}(T) =  \frac{F(T/\Tk)}{T} \, .  
 \end{align}
The meaning of \Eq{eq:ScalingFunction} is that  as long as 
the temperature is much smaller than the bandwidth, $T \ll W$,  the dependence of $\chi_\mathrm{imp}(T)$ on the model parameters $\Jk$ and $W$
enters only via the scale $\Tk$.  Still, this does not mean  that spin screening ``occurs at $\Tk$'', as is sometimes asserted in the literature.  Both the scale $\Tk$ \textit{and} the scaling function $F$ are needed to characterize the full crossover  from an unstable  high-temperature fixed point 
to a stable low-temperature fixed point.

Wilson computed the scaling function $F$ numerically using
his newly-developed numerical renormalization group
approach.  Fitting his numerical results, he found that
$\chi_\mathrm{imp}(T)$ is well described by the following
three functional forms, applicable for high, intermediate,
and low temperatures, respectively (cf.\ Eq.~(4.53) of Ref.~\cite{Hewson1993}):
\begin{subnumcases}{\chi_\mathrm{imp}(T) \simeq}
\label{eq:chiLargeT}
 \! \tfrac{1}{4T} \! \left[1 \! - \! \tfrac{1}{\ln(T/\Tk)}  
+ \tfrac{\ln [\ln (T/\Tk)]}{2 [\ln(T/\Tk)]^2}  \right. 
\hspace{-4cm}
\nonumber 
  \\ \left. \qquad + \;
  \mathcal{O} \! \left( \tfrac{1}{[\ln(T/\Tk)]^3}\right) \right]\! , 
 & $(T \! > \! T_2)$, 
 \\
 \label{eq:ChiIntermediateT}
 \! \tfrac{0.68}{4} \tfrac{1}{T + \sqrt{2} \Tk} , 
 & 
 $ (T_1 \! < \! T \! < \! T_2)$,  
\\
 \label{eq:chiLowT}
\! \tfrac{0.4132}{4 \Tk} \! \left[ 1 \!-\! 
\mathcal{O} \left(\tfrac{T}{\Tk}\right)^{\!2} 
\right] \! , 
& $ (T \! < \!  T_1)$.  \hspace{2mm} 
\end{subnumcases}
Several comments are in order. First, 
Wilson defined $\Tk$ via a high-temperature condition,
namely that the expansion \eqref{eq:chiLargeT} of $T\chi_\mathrm{imp}(T)$
should not contain a $[\ln(T/\Tk)]^{-2}$ term. 
Notice, however, that  the definition of $\Tk$ in terms of bare parameters is not unique, as it depends on the
cutoff procedure, as discussed by Wilson himself or
in Hewson's book \cite{Hewson1993}. 
Indeed, a  change in the definition of $\Tk$ can always be  compensated by   a change in the scaling function $F$. 

Second, $T_2$ and $T_1$ are the scales
where deviations from the high- or low-temperature
forms,  \eqref{eq:chiLargeT} or \eqref{eq:chiLowT},
first become noticeable when $T$ is decreased below $T_2$
or increased above $T_1$, respectively.  Their values depend on the definition of $\Tk$; for that of Wilson, they are given by $T_2 = 16 \Tk$ and $T_1 = 0.5 \Tk$
(see Eq.~(IX.99) in Ref.~\cite{Wilson1975} and Hewson \cite{Hewson1993}).
In the parlance of the main text of this paper,
they may be viewed as the onset and completion of spin screening scales, $\TspinO$ and $\TspinC$, respectively.

Third, we discuss the three functional forms given above.
The high-temperature fixed point describes an essentially free 
local moment. Correspondingly, the high-temperature susceptibility, \Eq{eq:chiLargeT}, shows Curie behavior $\chi_\mathrm{imp} \sim 1/T$ with logarithmic  corrections due to a marginally relevant operator. The crossover regime of intermediate
temperatures shows Curie-Weiss behavior, \Eq{eq:ChiIntermediateT}.
The overall prefactor, $0.68/4$, is about 30\% smaller than
the prefactor $1/4$ of the pure Curie law \eqref{eq:chiLargeT},
reflecting the renormalization of the impurity magnetization due
to the onset of screening with lowering temperature. 
The low-temperature fixed point describes FL excitations scattering off a fully screened impurity. 
Correspondingly, the low-temperature susceptibility, 
\Eq{eq:chiLowT},  approaches a constant for
$T/\Tk \to 0$, with a $(T/\Tk)^2$ correction caused by a leading irrelevant operator. The zero-temperature value of $4 \Tk \chi_\mathrm{imp}(0) = 0.4132$, known as the Wilson number, is a characteristic property
of the crossover function, linking properties
of the high-and low-temperature fixed points.

Fourth, we note  that an exact expression for the scaling
function $F$ was later obtained using the Bethe Ansatz,
\cite{Andrei1981,Filyov1981,Tsvelick1983}. In particular, Andrei and Lowenstein obtained an analytical expression for the Wilson number
\cite{Andrei1981}. The definitions of  $\Tk$ used in the Bethe Ansatz papers differ from that of Wilson, but the universal behavior of the susceptibility agrees with Wilson's solution.  The universality results from two facts: first, the impurity model is studied at very  weak coupling ($\Jk \ll W$), and second,
there is only one (marginally) relevant  operator perturbing the unstable fixed point \cite{Hewson1993}. 

Fifth, we note for completeness that Wilson's version of our
\Eq{eq:ChiIntermediateT}, namely his (IX.99), contains a 
factor 2 instead of $\sqrt{2}$ in the denominator. 
That is a typo, first noticed by Mel'nikov~\cite{Melnikov1982}, see 
p.~503 of Ref.~\cite{Tsvelick1983}, and also Ref.~\cite{Hewson1993}, below 
Eq.~(4.60).

To conclude our summary of Wilson's results on $\chi_\mathrm{imp}(T)$, we 
emphasize again that spin screening  is a gradual crossover phenomenon, even in the simple context of the Kondo impurity  model.  To 
describe the crossover quantitatively, it does not suffice to specify
just a single number for the crossover scale, even when only a single scale is dynamically generated.  Instead, one also has to specify 
which  observable and which scaling function was used,  and  
the precise criteria used to define the crossover scale.

Now let us discuss the relevance of the above arguments for
the present paper. 
DMFT maps 
the Hund-Hubbard lattice model that we consider in the main text 
to a quantum impurity model with a self-consistent bath.  The bath  is described by a hybridization function, which, in contrast to the pure Kondo model studied by Wilson, has a non-trivial structure. Moreover,
this structure depends on temperature.  Nevertheless  Wilson’s NRG approach for solving impurity models has been generalized to accommodate these complications, and indeed is now a widely-used impurity
solver for DMFT. 

Some  of  the terminology introduced by    Wilson and  reviewed above  can  also be used to understand some aspects of the solution of the DMFT equations and to illuminate the physics of the problem.
 For Hund metals, we have shown in Ref.~\cite{Stadler2015}
that an impurity with a rigid (not self-consistent) bath is a good guide to the full DMFT solution. Moreover, we argued there that Hund metals 
can be characterized by the criterion that the crossover
scales for spin and orbital screening differ strongly, 
$\Tspin \ll  \Torb$,  implying SOS, 
to identify a Hund metal.  In that work, as here,
we defined $\Tspin$ and $\Torb$ as the energy scales at which 
the imaginary parts of the zero-temperature dynamical spin and orbital susceptibilities are maximal. We emphasize, though, that the occurrence or not of spin-orbital separation does not depend on the criteria used to define these crossover scales.  For example, the onset-of-screening scales discussed in 
Sec.~\ref{sec:SusceptibilityH1} likewise yield $\TspinO \ll \TorbO$ for the Hund system \Hone{}.

In Ref.~\cite{Deng2019}, we refined our discussion
of crossover scales by analyzing the temperature dependence
of the spin and orbital susceptibilities, $\chi_\spin(T)$
and $\chi_\orb (T)$.  We  introduced  
onset-of-screening scales $\TspinO$, $\TorbO$ below
which deviations (say by $x_1 \%)$ from pure Curie behavior set in, 
and  completion-of-screening scales $\TspinC$, $\TorbC$
above which deviations (say by $x_2 \%)$ from pure Pauli behavior set in.
They correspond to Wilson's $T_2$ and $T_1$ scales, respectively.
These operational definitions have some degree
of arbitrariness (through the choices of $x_1$ and $x_2$;
in fact, these were not even specified in Ref.~\cite{Deng2019}). However, they have the advantage that they can also be applied 
when the crossover function in the intermediate temperature regime does not have a simple analytical form, a situation generally encountered for self-consistent DMFT impurity models. 
We argued in Ref.~\cite{Deng2019}
that the onset temperatures are useful to distinguish Mott systems from Hund systems: in Hund systems we have $\TspinO \ll \TorbO$, 
but in Mott systems  $\TspinO \simeq \TorbO$,
since the onset of spin and orbital screening 
with decreasing temperature occurs around the same temperature $\TM$ at which a quasiparticle peak begins to emerge
from the Mott pseudogap. Again, this distinction between Hund and Mott systems does not depend on  the precise criteria used to define the onset scales. 

In the main text of the present paper, we refined our
discussion of crossover scales somewhat more. 
We exploited the freedom in the choice of definition of the 
onset and completion scales (i.e., of $x_2$ and $x_1$) 
to reduce  the number of parameters by defining $\TorbC \simeq \TspinO$  for Hund systems and $\TorbC \simeq \TspinC$ for Mott systems. 
These choices, compatible with our data for \Hone\ and \Mone, 
have simple physical interpretations: For Hund systems,
featuring SOS, 
spin screening sets in once orbital screening is complete. By
contrast, for Mott systems, spin and orbital screening go hand in hand:
just as both onset-of-screening  scales
coincide with the emergence of a quasiparticle
peak from the Mott pseudogap and therefore 
match, $\TspinO \simeq \TorbO \simeq \TM$,
the completion-of-screening scales match, too, $\TorbC \simeq \TspinC$. 

\begin{figure}
\includegraphics[width=\linewidth]{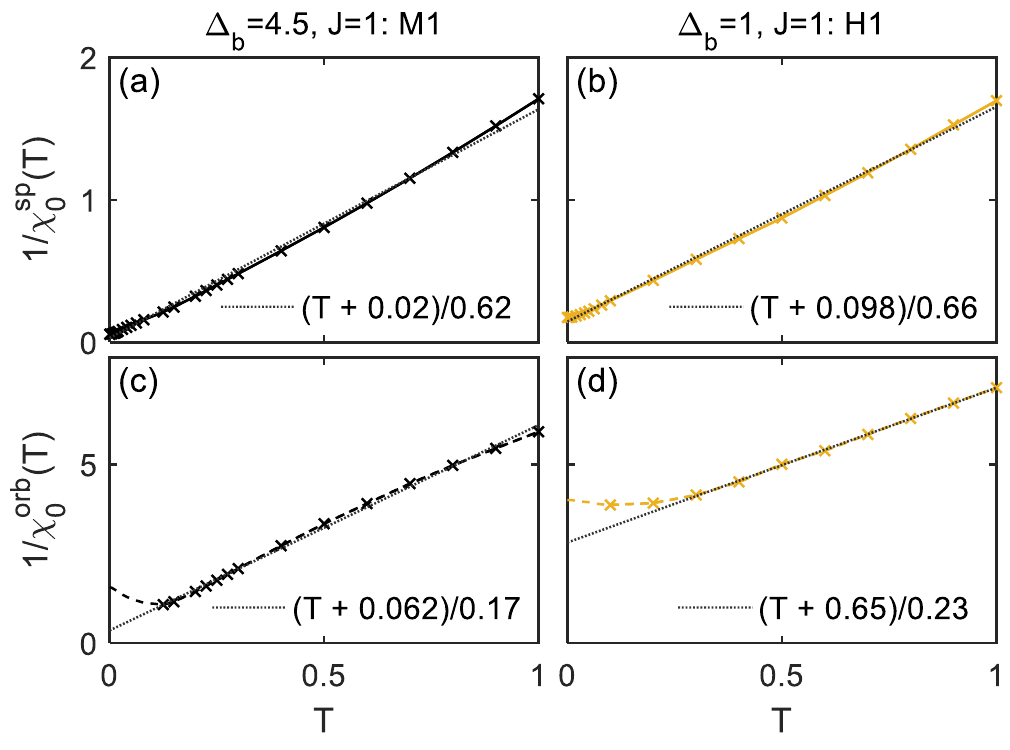}
\caption{Testing the applicability of a Curie-Weiss (CW) form
for various susceptibilities by replotting the data from
Fig.~\ref{fig:Tdep_1}(c) as $1/\chizero(T)$ vs $T$. 
The top row shows the spin susceptibilities
of M1 (left) and H1 (right) using solid lines, 
the bottom row the same for
the orbital susceptibilities, using dashed lines.
Dotted lines show Curie-Weiss fits to those 
data points (shown using crosses) for temperatures higher than the temperature at which $\chizero (T)$ is maximal.}
\label{fig:CW} 
\end{figure}

As a final remark, we note that one may attempt \cite{Mravlje2016,Katanin2021,Deng2021} to 
characterize the spin susceptibility $\chizero(T)$ of Hund systems using the  Curie-Weiss (CW) form $\chi^\CW_0 (T) = \mu /(T + \theta)$, with $\theta$
serving as a crossover scale. 
The CW form applies if a plot of $1/\chizero(T)$ vs $T$ yields a straight line. 
Figures~\ref{fig:CW}(a) and \ref{fig:CW}(b) show such plots for the spin susceptibilities of \Mone{} and \Hone{}. The resulting curves show clear deviations from linear behavior,
in particular for large $T$. Therefore, CW fits (dotted lines)  
characterize these susceptibilities only fairly crudely (see also 
Refs.~\cite{Katanin2021,Deng2021}). For completeness,
Figs.~\ref{fig:CW}(c) and \ref{fig:CW}(d) show analogous plots of the orbital susceptibilities.
These curves are strongly nonlinear  in the low-temperature regime
corresponding to the completion of orbital screening, where the CW form
is not applicable at all. 

\section{Temperature dependence of optical conductivity}
\label{sec:TdepQP}
\begin{figure*}
\centering
\includegraphics[width=0.9\linewidth, trim=14mm 175mm 14mm 0mm, clip=true]{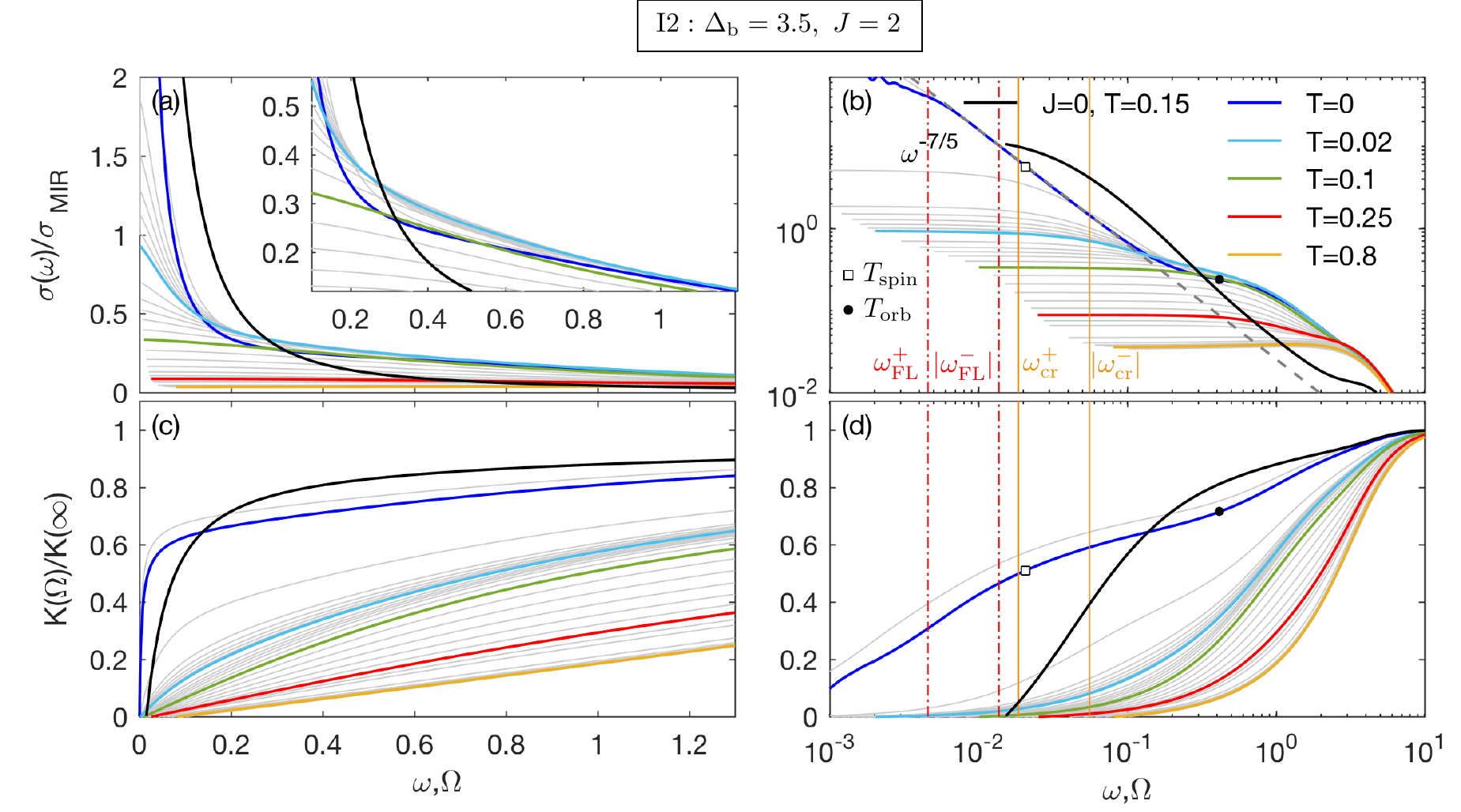}
\caption{[(a),(b)] The optical conductivity $\sigma(\omega)$ and the
  kinetic energy $K(\Omega)$ are plotted for various temperatures on
  [(a),(c)] a linear and [(b),(d)] a logarithmic frequency scale for \Itwo{} ($\Delta_b=3.5$, $J=2$). 
In addition, data for \Wzero{} ($\Delta_b=3.5$, $J=0$) at $T=0.15$ is shown in black.  [(b),(d)] $|\omega^{\pm}_{\fl}|$,
  below which FL behavior should set in, is marked by vertical
  dash-dotted red lines.  The vertical solid yellow lines denote $|\omega^{\pm}_\cross|$. 
  Filled dots and open squares mark the orbital and spin Kondo scales, respectively.}
\label{fig:asym_2}
\end{figure*}
We next study the optical conductivity $\sigma(\omega)$
[cf.\ \Eq{eq:optcond}]
again for system \Itwo. $\sigma(\omega)$ is plotted on a linear
and a logarithmic frequency scale in \Figs{fig:asym_2}(a) and \ref{fig:asym_2}(b),
respectively. For comparison, we also show data for \Wzero\
($\Delta_b=3.5$, $J=0$), computed at $T=0.15$, which is still in the
FL temperature regime. At $T=0$ we expect a FL Drude peak for \Itwo. However,
the data (cf.\ blue curve) is not accurate enough to resolve the FL
behavior at very low frequencies, $\omega<\omega^{+}_{\fl}$ 
(cf.\ discussion of blue and red curves in Fig.~3.1(b) in Sec.~3.2 of \oRef{Stadler2019}).
In the low-frequency NFL crossover regime,
here approximately given by $\omega^{+}_{\fl}\lesssim\omega\lesssim|\omega^{-}_\cross|$, we observe a power-law
flank in $\sigma(\omega) \propto\omega^{-\alpha}$, with
$\alpha\approx7/5$ at $T=0$. Notably, for $\omega>|\omega^{-}_\cross|$ a broad HQP shoulder develops around $\Torb$ at $T=0$. 
 
 With increasing temperature but below $T\lesssim\TspinO=0.1$, 
 spin degrees of freedom are gradually unscreened in the system 
 while the orbitals are still screened. This process is reflected in $\sigma(\omega)$:
with increasing temperature spectral weight is shifted from low frequencies into the HQP shoulder,
while the high-frequency flank of $\sigma(\omega)$ remains
unaffected. Note that the HQP shoulder is absent for $J=0$ [cf.\ black
  curve in \Figs{fig:asym_2}(a) and \ref{fig:asym_2}(b)]. At higher temperatures ($T>\TspinO$) the HQP
shoulder gradually decreases in height, reflecting the unscreening of the orbital degrees of freedom in \Itwo. The second shoulder at 
bare energy scales is a Hubbard-band feature,
which is also present for $J=0$. We suspect that
the HQP shoulder at $\omega>|\omega^{-}_\cross|$ is an 
optical fingerprint of the HQP band [$\textrm{SU(3)}$ 
Kondo resonance in $A(\omega)$] and can indeed be interpreted as Hund's-coupling-induced excess spectral
weight, caused by resilient QPs, as suggested in \oRef{Stricker2014}. Further, we remark that
our results (for $T\lesssim\TspinO$) are reminiscent of recent optical
conductivity measurements~\cite{Yang2017} for KFe$_2$As$_2$.

In \Figs{fig:asym_2}(c) and \ref{fig:asym_2}(d) the kinetic energy $K(\Omega)$ [as defined in \Eq{eq:kinen}] is plotted as a function of frequency $\Omega$ for various temperatures. In \oRef{Schafgans2012} an unusual spectral weight transfer from low to high energies was observed at low temperatures in $K(\Omega)$ for iron pnictides. This observation would correspond to  line crossings of different $K(\Omega,T)$ curves for $J=2$  in \Figs{fig:asym_2}(c) and \ref{fig:asym_2}(d), which is yet  not found in our data. We remark that this might be due to the rather large $\Delta_b=3.5$.

\section{Elementary definitions and relations}
\label{app:definitions}

\subsection{Optical conductivity, kinetic energy, resistivity, and the Mott-Ioffe-Regel (MIR) limit}
\label{sec:calc_optcond}
\paragraph*{Optical conductivity.}
The (real part of the) optical conductivity (per spinful band), 
computed in linear response, is given by \oRef{Deng2013},
\begin{align}
\label{eq:optcond}
\sigma(\omega) {}=&\frac{2\pi e^2}{\hbar}  \int {\text d}\omega'\, \frac{f(\omega')-f(\omega+\omega')}{\omega} \nonumber 
\\
& \times \int {\text d}\varepsilon\,\Phi(\varepsilon)A(\varepsilon,\omega')A(\varepsilon,\omega+\omega'), 
\end{align}
where $f(\omega)$ is the Fermi function, 
$A(\varepsilon,\omega)$ the structure factor as defined in \Eq{eq:ARPES},
and $\Phi(\varepsilon)$ the transport velocity kernel,
\begin{subequations}
 \label{eq:Phi}
 \begin{align}
\Phi(\varepsilon)
& =  \int \frac{\textrm{d}^d k}{(2\pi)^d}\,
\left(\frac{\partial\varepsilon_{\mathbf{k}}}{\partial k_x}\right)^2\delta(\varepsilon-\varepsilon_{\mathbf{k}}) 
\\
& = 
\Phi(0)\left[1-\left(\frac{\varepsilon}{D}\right)^2\right]^{\frac{3}{2}}.  \label{eq:PhiBethe}
 \end{align}
\end{subequations}
The latter is here expressed through the
 band velocity in $x$ direction, 
$v_{\mathbf{k}}^x= \frac{\partial\varepsilon_{\mathbf{k}}}{\hbar \partial k_x}$, 
and \Eq{eq:PhiBethe} follows for the Bethe lattice.

\paragraph*{Kinetic energy.}
The kinetic energy $K(\Omega)$ is the integral of the optical conductivity 
$\sigma(\omega)$ up to a cutoff value $\Omega$ \cite{Schafgans2012}:
\begin{align}
\label{eq:kinen}
\frac{K(\Omega)}{K(\infty)}=
\frac{\int_0^{\Omega}{\textrm d}\omega\,\sigma(\omega)}
{\int_0^{\infty}{\textrm d}\omega\,\sigma(\omega)}.
\end{align}
We normalize $K(\Omega)$ to $K(\infty)$.

\paragraph*{Resistivity.}
\label{sec:appendix_res}
The temperature-dependent optical resistivity is given as the inverse of the optical conductivity 
evaluated at the Fermi level, $\omega=0$, 
\begin{equation}
\rho(T)=\frac{1}{\sigma(\omega=0,T)}.
\end{equation}
\paragraph*{Mott-Ioffe-Regel (MIR) limit.}
In conventional  metals $\rho(T)$ increases with temperature. This behavior can be explained in a QP picture: 
the mean-free path $l$ of a QP gradually decreases because thermally-induced scattering events 
become more frequent. For phonon scattering at higher temperatures, i.e., above a small temperature 
below which electron-electron  scattering  is dominant, this leads to a linear growth of $\rho(T)\sim T$.
However, this QP picture breaks down approximately when $l$ becomes shorter than the interatomic spacing, 
leading to the Mott-Ioffe-Regel (MIR) limit,
$k_{\textrm{F}}l_{\textrm{min}}\approx2\pi$ \cite{Ioffe1960,Mott1972,Gurvitch1981}
(another popular definition is $k_{\textrm{F}}l_{\textrm{min}}\approx1$).
As a consequence, above a corresponding MIR temperature $T_\MIR$, the resistivity saturates in conventional metals, approaching a maximum value
$\rho_\MIR$. While for most good metals, $l\gg 2 \pi / k_{\textrm{F}}$ holds up to their melting temperatures, there is a vast number of 
metals for which the MIR resistivity saturation is observed \cite{Hussey2004}.
Interestingly, most strongly correlated metals, like cuprate high-temperature superconductors (HTSCs), heavy fermions, Hund metals (including 
iron-based HTSCs), and also several organic compounds exceed the MIR limit and $\rho(T)$ does \textit{not} saturate with increasing temperature.
Due to this unconventional but common feature, which is generically assumed to be induced by some kind of NFL behavior, 
all these materials are collectively referred to as ``bad metals'' in the literature \cite{Emery1995,Hussey2004}.

In \Fig{fig:asym_2} $\sigma(\omega)$ 
is measured in units of $\sigma_\MIR=\frac{2\pi e^2\Phi(0)}{\hbar D}$.
This is the MIR  limit  derived in \oRef{Deng2013} 
for a free parabolic
band in two dimensions, $\varepsilon(\mathbf{k})=
\tfrac{\hbar^2(k_x^2+k_y^2)}{2m}$, 
using the criterion $k_{\text{\fl}}l_{\text{min}}= 2\pi$. 
Accordingly, in Fig.~\ref{fig:Tdep_2} we plot $\rho$ in units of $\rho_\MIR
=1/\sigma_\MIR$.

\subsection{Thermopower}
\label{sec:thermopower}
The thermopower (Seebeck coefficient)
is defined as $\alpha(T)=-\Delta \mathcal{V}/\Delta T$, 
where $-\Delta \mathcal{V}$ is the electric field 
generated when a thermal gradient $\Delta T$ is established
in a material under conditions which are such that no electrical current flows \cite{Mravlje2016}.
We calculate $\alpha(T)$  with the Kubo formula of \oRef{Mravlje2016},
\begin{equation}
\label{eq:alpha}
  \alpha(T)  = -\frac{k_B}{e} \frac{\int {\textrm{d}}\omega \,T(\omega)\beta \omega \left(-\frac{\partial f}{\partial \omega}\right)}{\int {\textrm{d}}\omega \,T(\omega)\left(-\frac{\partial f}{\partial \omega}\right)}, 
\end{equation}
where $\beta=1/k_{\textrm{B}}T$,
and the transport function $T(\omega)$ given here for transport in
$x$ direction, reads
\begin{align}
T(\omega) &= 
2\pi e^2 \int \frac{\textrm{d}^d k}{(2\pi)^d}\,
\left(v_{\mathbf{k}}^x\right)^2 A_{\mathbf{k}}(\omega)^2 \nonumber \\
\label{eq:T}
& = \frac{2\pi e^2}{\hbar^2} 
\int {\textrm{d}} \varepsilon\, \Phi(\varepsilon) A(\varepsilon,\omega)^2 .
\end{align}

\subsection{Entropy}
\label{sec:calc_entropy}
Within DMFT, where a lattice system is mapped self-consistently onto an impurity system, we can both calculate the impurity contribution to the entropy, as usually done within NRG~\cite{Bulla2008}, and the lattice entropy. Importantly, these entropies differ (quantitatively but not qualitatively), as is discussed in detail in \Sec{sec:entropy}. 

\paragraph*{Impurity contribution.}
The impurity contribution to the entropy $S_\imp$ is introduced in Eqs.~(48) and (53) of \oRef{Bulla2008}
as the difference,
\begin{equation}
\label{eq:Simp}
S_\imp(T)=S_{\textrm{tot}}(T)-S_{\textrm{tot}}^{(0)}(T),
\end{equation}
between the entropy of the total Wilson chain $S_{\textrm{tot}}$ and the entropy of a reference system $S_{\textrm{tot}}^{(0)}$, which is the bare conduction Hamiltonian without impurity. In practice, it is thus necessary to perform two independent NRG runs, one for the full Hamiltonian and one for the same Hamiltonian without impurity.

\paragraph*{Lattice entropy.}
Starting from the thermodynamic relation 
$T (\partial S_\latt /\partial T) = \partial \mathcal{E}_\latt/ \partial T$ between the entropy and the total internal
energy of the lattice, the lattice entropy can be expressed
as an integral involving the specific heat,  
$C(T)=(\partial\mathcal{E}_\latt/ \partial T)$,
\begin{equation}
\label{eq:Slatt}
S_{\latt}(T)=S_\latt(T_0)+\int_{T_0}^T{\rm d} T'\, \frac{C(T')}{T'} \, , 
\end{equation}
following Eq.~(238) of \oRef{Georges1996}.  $S_\latt(T_0)$ is a
constant offset, in principle unknown. In the case of a FL, however,
$S_\latt(T_0)$ can be determined exactly [cf.\ \Eq{eq:entropy}].
For Hubbard-type models in the limit of large
lattice coordination, the total internal energy is given by Eq.~(7) of
\oRef{Kotliar1999}, which we apply in the form,
\begin{subequations}
\begin{align}
\frac{\mathcal{E}_\latt}{N_c} {}
= & \int{\rm d} \omega\, f(\omega)(\omega+\mu)A(\omega) 
\\
& +2t^2\int{\rm d}\omega_1 \int{\rm d}\omega_2 \, f(\omega_1)\frac{A(\omega_1)A(\omega_2)}{\omega_1-\omega_2} 
\nonumber \\
\label{eq:Etot2}  =&
\int{\rm d} \omega\, f(\omega)(\omega+\mu)A(\omega)\\
& -\frac{2t^2}{\pi}\int{\rm d}\omega \, f(\omega)\real G(\omega) \imag G(\omega). \nonumber 
\end{align}
\end{subequations}
Here $f(\omega)$ is the Fermi function,
and the second equality follows via 
the Kramers-Kronig relation, $\real G(\omega)=\tfrac{1}{\pi} P \int{\rm d} \omega' \, \tfrac{\imag G(\omega')}{\omega'-\omega}=P \int {\rm d} \omega' \, \frac{A(\omega')}{\omega-\omega'}$.

%

\end{document}